\documentclass[final,3p,times,fleqn]{elsarticle}

\usepackage{amssymb,amsthm}

\journal{Physica A}

\usepackage[dvipdfmx]{hyperref}
\usepackage{xcolor}
\usepackage{textcase}
\usepackage{amsmath,amssymb,amsfonts}
\usepackage{mathrsfs}
\usepackage{mathtools}
\usepackage{braket}
\usepackage{enumitem} 
\usepackage{booktabs,arydshln}
\usepackage{threeparttable}
\biboptions{sort&compress}
\usepackage{empheq}

\usepackage[font={small},labelsep=space]{caption}
\DeclareCaptionLabelFormat{tablelab}{\textbf{Table\,{#2}.}}
\DeclareCaptionLabelFormat{figurelab}{\textbf{Fig.\,{#2}.}}
\newcommand{\f}[2]{\frac{#1}{#2}}
\newcommand{\hf}[2]{\text{\small $\frac{#1}{#2}$}}
\newcommand{\ko}[1]{\left( #1 \right)}
\newcommand{\kko}[1]{\left[ #1 \right]}

\newcommand{\abs}[1]{\left| #1 \right|}
\newcommand{\q}[1]{`#1'}
\newcommand{\bs}[1]{\boldsymbol{#1}}
\newcommand{\vev}[1]{\left\langle{#1}\right\rangle}

\DeclareMathOperator{\diag}{diag}

\def\iD{\mathit{\Delta}}
\def\iS{\mathit{\Sigma}}
\def\ds{\displaystyle}
\def\pa{\partial}
\def\eq{\equiv}
\def\cS{\mathcal{S}}
\def\cN{\mathcal{N}}
\def\cL{\mathcal{L}}
\def\cH{\mathcal{H}}
\def\cM{\mathcal{M}}
\def\cU{\mathcal{U}}
\def\cV{\mathcal{V}}
\def\cZ{\mathcal{Z}}

\def\cB{\textsf{B}}
\def\ep{{\epsilon}}
\def\no{\nonumber}

\begin{document}

\begin{frontmatter}

\title{Affinity-based extension of non-extensive entropy and statistical mechanics}
\author{Keisuke Okamura\corref{cor1}}
\address{Ministry of Education, Culture, Sports, Science and Technology, 3-2-2 Kasumigaseki, Chiyoda-ku, Tokyo 100-8959, Japan.}
\ead{okamura@ifi.u-tokyo.ac.jp}

\begin{abstract}
Tsallis' non-extensive entropy is extended to incorporate the dependence on affinities between the microstates of a system.
At the core of our construction of the extended entropy ($\cH$) is the concept of the effective number of dissimilar states, termed the \q{effective diversity} ($\iD$).
It is a unique integrated measure derived from the probability distribution among states and the affinities between states.
The effective diversity is related to the extended entropy through the Boltzmann's-equation-like relation, $\cH=\ln_{q}\iD$, in terms of the Tsallis' $q$-logarithm.
A new principle called the Nesting Principle is established, stating that the effective diversity remains invariant under an arbitrary grouping of the constituent states.
It is shown that this invariance property holds only for $q=2$; however, the invariance is recovered for general $q$ in the zero-affinity limit (i.e.\ the Tsallis and Boltzmann--Gibbs case).
Using the affinity-based extended Tsallis entropy, the microcanonical and the canonical ensembles are constructed in the presence of general between-state affinities.
It is shown that the classic postulate of equal {a priori} probabilities no longer holds but is modified by affinity-dependent terms.
As an illustration, a two-level system is investigated by the extended canonical method, which manifests that the thermal behaviours of the thermodynamic quantities at equilibrium are affected by the between-state affinity.
Furthermore, some applications and implications of the affinity-based extended diversity/entropy for information theory and biodiversity theory are addressed in appendices.
\end{abstract}

%


\end{frontmatter}


\section{Introduction\label{sec:level1}}
In classical statistical mechanics, Boltzmann--Gibbs entropy \cite{Gibbs02}, given by
\begin{equation}\label{Shannon}
H_{1}(\{p_{i}\})=-\sum_{i\in\cS} p_{i}\ln p_{i}\,,
\end{equation}
plays a central role to interrelate various thermodynamic quantities.\footnote{The Boltzmann constant $k_\mathrm{B}$ is set to be unity throughout this paper.}
Here, $\cS\coloneqq\{1,\,\dots,\,n\}$ 
represents the set of labels for microstates that are consistent with the given macrostate of a system, and $p_{i}$ denotes the probability of realising the $i$-th microstate, satisfying $p_{i}\in [0,1]$ and $\sum_{i\in\cS} p_{i}=1$.
An important property of this entropy function is that it reaches its maximum for the uniform probability distribution, 
\begin{equation}\label{Boltzmann}
H_{1}^{\max}=H_{1}(\{p_{i}=1/n\})=\ln n\,,
\end{equation}
which is the celebrated Boltzmann's equation.
The entropy function (\ref{Shannon}) has mathematically the same form as Shannon entropy \cite{Shannon48} in information theory, and its possible extension has been widely investigated for its potential applications in various contexts \cite{Renyi61,Havrda67,Daroczy70,Patil82,Tsallis88,Tsallis99,Jizba04,Tsallis09,Leinster12}.
One of the most well studied amongst these is Tsallis entropy \cite{Tsallis88,Tsallis99,Tsallis09,Abe-Okamoto01}, defined by
\begin{equation}\label{Tsallis}
H_{q}(\{p_{i}\})
\coloneqq -\sum_{i\in\cS}(p_{i})^{q}\ln_{q}p_{i}
=\f{1}{1-q}\Bigg(\sum_{i\in\cS}(p_{i})^{q}-1\Bigg)\,,
\end{equation}
where the Tsallis' $q$-logarithmic function is defined by $\ln_{q} x\coloneqq (x^{1-q}-1)/(1-q)$ with $q\in [0,\infty)\backslash \{1\}$.
This extended entropy function is undefined for $q=1$, but its limit as $q$ approaches $1$ is well-defined and reproduces Boltzmann--Gibbs entropy: $\lim_{q\to 1}H_{q}=H_{1}$ as $\ln_{q} x\to \ln x$.
Subsequently, what we term the \emph{diversity} of a system, denoted as ${}^{q}\!D$, is defined through the following Boltzmann's-equation-like relation,
\begin{equation}\label{H=ln_qD}
H_{q}(\{p_{i}\})=\ln_{q}{}^{q}\!D(\{p_{i}\})\,,
\end{equation}
which is solved to give the diversity index
\begin{equation}\label{Hill_num}
{}^{q}\!D(\{p_{i}\})
=\begin{cases}
\, \ds \Bigg(\sum_{i\in\cS} (p_{i})^{q}\Bigg)^{\f{1}{1-q}} & (q\neq 1) \\[4mm]
\, \ds \exp\Bigg(-\sum_{i\in\cS} p_{i}\ln p_{i}\Bigg) & (q=1)
\end{cases}\,.
\end{equation}
This essentially represents the so-called entropy power of order $q$ (also known as the Uffink's class of entropies \cite{Uffink95,Jizba19}) as originally introduced by Shannon \cite{Shannon48} for the $q=1$ case.
Interestingly, this measure has also been known in the field of quantitative biology for close to a half-century with the name of Hill numbers \cite{Hill73}.
It quantifies the diversity of an arbitrary system under a special condition that there is no similarity between any pair of the involved \q{categories}, here labelled by $i\in\cS$.
Then, the diversity depends \q{exogenously} on only a single aspect, which is the evenness of the element distribution among categories characterised by $\{p_{i}\}$.
In addition, the diversity is also influenced by an \q{endogenous} aspect, i.e.\ the parameter $q$ determining how much weight to place on the category prevalence.

These are, however, necessary but individually insufficient aspects of what constitutes diversity in general. 
As is well-acknowledged \cite{Rao82,Leinster12}, essential defining characteristics of diversity include an additional exogenous aspect, i.e.\ the similarity between categories, which the Hill numbers family fails to take into account.
There also arises an additional endogenous aspect of diversity, i.e.\ how much weight to place on between-category dissimilarity, which has attained little attention so far.
Thus, these two exogenous plus two endogenous aspects are essential to properly characterise and quantify diversity.\footnote{In biological contexts, e.g.\ species diversity, it is common to characterise diversity by three aspects; namely, species richness (or \q{variety}; $n$), distribution across species (or \q{balance}; $\{p_{i}\}$) and \q{dissimilarity} ($\{d_{ij}\}$) between species.
In relation to the current terminology, the \q{balance} and \q{dissimilarity} aspects correspond to the two exogenous aspects, while the \q{variety} aspect can be incorporated in the definition of the \q{balance} aspect as $n\coloneqq \abs{\{i\in\cS\,|\,p_{i}>0\}}$.\label{fnlabel}}

In statistical mechanics, these two exogenous aspects of diversity correspond to the evenness of probability distribution among microstates and the degrees of similarity, or affinity, between microstates.
In addition, the two endogenous aspects correspond to how much weight to place on the probability of realising individual states and how much weight to place on the between-state affinities.
Tsallis and Boltzmann--Gibbs entropies can be regarded as corresponding to situations in which all the microstates are \q{maximally distinct}, meaning that all the between-state affinities are zero.
Then, we are naturally motivated to investigate the implications of the new degrees of freedom---the between-state affinities---on the notion of entropy.
It is intriguing to see whether such an extended entropy, obtained from the affinity-based extended diversity index, provides a useful tool to further probe and understand the physics of statistical mechanics and other areas of scientific disciplines.
In fact, in some disciplines, the notion of affinity-based diversity/entropy naturally emerges and exists in reality (see, e.g., \ref{app:biodiversity} for the biodiversity context).
With this motivation, in this paper, we propose a possible extension of Tsallis entropy in this direction and demonstrate its implications for the non-extensive statistical mechanics with some concrete examples.

The rest of this paper is organised as follows.
Section \ref{sec:2} provides a unified framework for quantifying entropy that incorporates all of the aforementioned essential aspects of diversity, including the possible between-state affinities.
Subsequently, a new principle called the Nesting Principle is established, stating that diversity/entropy remains invariant under an arbitrary grouping (i.e.\ an arbitrary nesting or partitioning) of the constituent categories or individual elements.
The formula for the effective diversity/entropy is derived along with the formula for the effective affinity between systems.
Section \ref{subsec:non-extensive} discusses some implications of the affinity-based extended Tsallis entropy on the non-extensive statistical mechanics.
Finally, Section \ref{sec:summary} is devoted to the summary and discussions.
\ref{app:diversity indices} summarises common diversity indices/entropies known in the literature and their relation to the present study.
\ref{app:divergence} presents some information-theoretic properties of the affinity-based extended Tsallis type entropies for future applications.
\ref{app:biodiversity} addresses the issue of diversity partitioning in a general biological context.

\section{Affinity-Based Extension of Tsallis Entropy\label{sec:2}}

In this section, we derive a general formula for the affinity-based extended Tsallis entropy.
The derivation is based on the foundation of the Hill numbers family (\ref{Hill_num}) with a new set of axioms and postulates.

\subsection{Preliminaries\label{subsec:preliminaries}}

Let us consider an assemblage (\q{a system}) in which individual elements (\q{point probabilities}) are distributed over $n$ categories (\q{microstates}).
Suppose that there are no between-category similarities.
Then, the only feature that influences the diversity of the assemblage is the distribution of elements over the categories.
As before, let $p_{i}$ denote the detection probability of the $i$-th category in any randomly observed element such that $p_{i}\in [0,1]$ and $\sum_{i\in\cS} p_{i}=1$ with $i\in \cS=\{1,\,\dots,\,n\}$.
Our first problem is how to define the diversity index depending exogenously only on $\{p_{i}\}$,
\begin{equation}
D=D(\{p_{i}\})\,,
\end{equation}
as a measure to quantify the diversity of an arbitrary assemblage, which can also be used to compare the diversity between two arbitrary assemblages.
The key idea to solve this problem is to \q{normalise} the assemblage such that it can be regarded as composed of distinct categories with the same relative abundances, thereby the resulting category richness serves as a unique indicator of diversity.
This normalisation procedure can be achieved as follows.

Suppose that some observable quantity $x$ is defined and measured for each category $i\in \cS$, giving a set of $n$ positive real numbers, $\{x_{i}\}$.
An \emph{effective} quantity can then be defined from the observed raw quantities via an averaging procedure.
A fairly general such procedure is the so-called weighted power mean (see, e.g., \cite{Hardy52}), defined for $\{x_{i}\}$ as
\begin{equation}\label{power_mean}
M_{\alpha}\big[\{x_{i}\} | \{w_{i}\}\big]\coloneqq \Bigg(\sum_{i\in \cS}w_{i}x_{i}{}^{\alpha}\Bigg)^{1/\alpha}\,,
\end{equation}
where $\{w_{i}\}$ is a sequence of non-negative weights which sum to unity, $\sum_{i\in\cS}w_{i}=1$, and $\alpha$ is a non-zero real parameter that controls the mean's sensitivity to the magnitude of $x_{i}$.\footnote{Special cases with equal-weights ($w_{i}=1/n$) include the arithmetic means ($M_{1}$), the geometric means ($\lim_{\alpha\to 0}M_{\alpha}$) and the harmonic means ($M_{-1}$).
In addition, the two extremals $\lim_{\alpha\to +\infty}M_{\alpha}$ and $\lim_{\alpha\to -\infty}M_{\alpha}$ give the maximum and the minimum of $\{x_{i}\}$, respectively.}
Let us also introduce its integral analogue defined by
\begin{equation}\label{power_mean_int}
\cM_{\alpha}\big[x | w\big]\coloneqq \Bigg(\int_{\cS}w(t)x(t){}^{\alpha}dt\Bigg)^{1/\alpha}\,,
\end{equation}
where $t$ parameterises the support of the distribution ($\cS$) such that the weight function, $w(t)$, satisfies $\int_{\cS}w(t)dt=1$.
In the normalised situation, an equal relative abundance for the categories (i.e.\ a constant weight function) implies that the effective diversity is given by the volume of the support, i.e.\ $D=\int_{\cS}dt$.
Intuitively, this can be regarded as a hypothetical assemblage containing $D$ equiprevalent categories, only this time with $D$ a real number rather than an integer.

Now, set the nominal weight $w_{i}$ in (\ref{power_mean}) to be equal to the probability $p_{i}$, and set the weight function $w$ in (\ref{power_mean_int}) to be equal to the constant weight, which is given by $1/D$.
The condition for $D$ to represent the desired effective diversity is that these two quantities---the weighted power means for the observed state and that for the normalised equivalent---agree with one another, i.e.\
\begin{equation}\label{Hill_eq}
M_{q-1}\big[\{p_{i}\}|\{p_{i}\}\big]
\eq \cM_{q-1}\kko{\hf{1}{D} \Big| \hf{1}{D}}\,.
\end{equation}
Here, the power parameter has been set as $q-1\eq\alpha$ by convention.
By solving (\ref{Hill_eq}) for $D$, we obtain exactly the Hill numbers family given in (\ref{Hill_num}), representing the effective number of categories \cite{Hill73,Jost06} such that $D\in [1,n]$.
It is conceptualised as the number of equally prevalent and maximally distinct categories that would yield the same diversity as that of the observed assemblage.

The diversity spectrum is continuous not only for the distribution $\{p_{i}\}$ but also in the parameter $q$, which is often referred to as the order parameter, indicating the diversity measure's sensitivity to common and rare categories.
Some specific values of $q$ are associated with well-known diversity/entropy metrics; 
see Table \ref{tab:Hill_num} of \ref{app:diversity indices} for summary.
If $q=0$, the index ${}^{0}\!D$ gives as much weight to rare category as common ones, simply counting the number of observed categories (\q{category richness}) \cite{MacArthur65};
if $q\in(0,1)$, the index assigns greater weight on relatively rare categories;
if $q=1$, it gives a \q{flat} weight, corresponding to the Boltzmann--Gibbs case \cite{Gibbs02}, $\ln{}^{1}\!D=-\sum_{i\in\cS} p_{i}\ln p_{i}$;
and if $q\in(1,\infty)$, it assigns greater weight on relatively common categories, with the limiting case $q\to\infty$ reflecting only the prevalence of the most common category: ${}^{\infty}\!D=1\big/\max_{i\in\cS}\{p_{i}\}$ \cite{Berger70}.
In particular, the $q=2$ case, ${}^{2}\!D=1\big/\sum_{i\in\cS}(p_{i})^{2}$, is related to the Gini--Simpson index and the Simpson concentration index \cite{Simpson49,Jost06}.

\subsection{Affinity-based extension\label{subsec:gen.formula}}

The Hill numbers family (\ref{Hill_num}) provided a measure of diversity incorporating the aspect of distribution evenness; however, the other exogenous aspect of between-category dissimilarity was ignored.
Given that our world is so complex that some categories can be similar to each other, while others can be very distinct from any others, it is natural to pursue a more general or realistic diversity formula that also incorporates the between-category dissimilarity aspect.
Mathematically, this amounts to obtaining a diversity index $\iD$---using a distinct notation from $D$ of Hill numbers---of the form
\begin{equation}\label{general_iD}
\iD=\iD(\{p_{i}\},\{d_{ij}\})\,,
\end{equation}
which not only depends on the distribution, $\{p_{i}\,|\,i\in\cS\}$, but also on the dissimilarities, $\{d_{ij}\,|\,i,\,j\in\cS\}$, where $d_{ij}$ denotes the observed dissimilarity between two categories represented by $i$ and $j$.
Here we take the convention that $d_{ij}\in [0,1]$ with $d_{ij}=0$ indicating identicality between $i$ and $j$, and $d_{ij}=1$ indicating complete dissimilarity between $i$ and $j$; by definition, $d_{ij}=d_{ji}$ and $d_{ii}=0$.
In addition, these dissimilarity variables are assumed to obey the triangle inequality, i.e.\ $d_{ij}+d_{jk}\geq d_{ik}$ for arbitrary triplets of categories $(i,\,j,\,k)$.

The formula of the form (\ref{general_iD}) integrates the effects of all the aforementioned essential aspects of diversity into a single dimension in the same unit.
Clearly, such a diversity index cannot be arbitrary, but it is necessarily constrained by the nature of the measure itself.
Given that, in contrast to the previous works (e.g.\ \cite{Leinster12}), this work is based on an axiomatic approach, i.e.\ we postulate the following Conditions \ref{cond:conti}--\ref{cond:even_p} as necessary conditions for any consistent dissimilarity-based diversity index:

\begin{enumerate}[labelindent=\parindent, label=\Roman*, align=parleft, leftmargin=!]
\item \textit{Continuity:} The diversity index is continuous in both the relative abundances of categories, $\{p_{i}\}$, and the between-category dissimilarities, $\{d_{ij}\}$.
\label{cond:conti}
\item \textit{Consistency (with the zero-affinity case):} If dissimilarity is set to be maximal for all distinct pairs of categories, i.e.\ $d_{ij}=1-\delta_{ij}$ with $\delta_{ij}$ Kronecker's delta, then the diversity index reduces to that derived from the equivalence relation (\ref{Hill_eq}) (i.e.\ the Hill numbers family).
\label{cond:reduceHill}
\item \textit{Invariance:} If two categories labelled by \q{1} and \q{2} are (in fact) identical (i.e.\ $d_{12}=0$), then the diversity index computed from the following two pictures are equivalent: (i) treating them as two separate entities (hence there are formally $n$ categories in total), and (ii) treating them as a single entity labelled by \q{0} (hence there are $n-1=|\{0,\,3,\,\dots,\,n\}|$ categories in total).\footnote{This condition can be viewed as a special case of the Nesting Principle introduced later in Section \ref{sec:NP}.\label{fn:nesting}}
\label{cond:split/merg}
\item \textit{Monotonicity:} If all constituent elements are uniformly distributed among $n$ categories (i.e.\ $p_{i}=1/n$ for all $i$) and the between-category dissimilarity is constant across all pairs of categories (i.e.\ $d_{ij}=\bar{d}$ for all $i\neq j$), then the diversity index is a monotonically increasing function of both the category richness $n$ and the constant dissimilarity $\bar{d}$.
\label{cond:even_p}
\end{enumerate}

\noindent
We use these four conditions as guiding principles to construct the dissimilarity-based diversity index.
It will be shown that the resulting diversity index also satisfies the following properties (as expected):

\begin{enumerate}[resume, labelindent=\parindent, label=\Roman*, align=parleft, leftmargin=!]
\item \textit{Maximality:} The diversity index is maximised to give the value $n$ if elements are uniformly distributed among the categories that are maximally dissimilar from one another:%
\label{cond:max}
\begin{align}\label{D_max}
n=\max_{\{p_{i}\},\{d_{ij}\}}\iD=\iD(\{\text{$p_{i}=1/n$ \,and\,  $d_{ij}=1-\delta_{ij}$}\})\,.
\end{align}
\item \textit{Minimality:} The diversity index is minimised to give the value $1$ if elements are distributed only among those categories with no dissimilarities, including the case of a single-category concentration (i.e.\ $p_{i}=1$ for some $i$):%
\label{cond:min}
\begin{align}\label{D_min}
1=\min_{\{p_{i}\},\{d_{ij}\}}\iD=\iD(\{\text{$d_{ij} =0$ \,if\,  $p_{i}p_{j}>0$}\})\,.
\end{align}
\end{enumerate}

\begin{figure}
\centering
\includegraphics[width=11.5cm,clip]{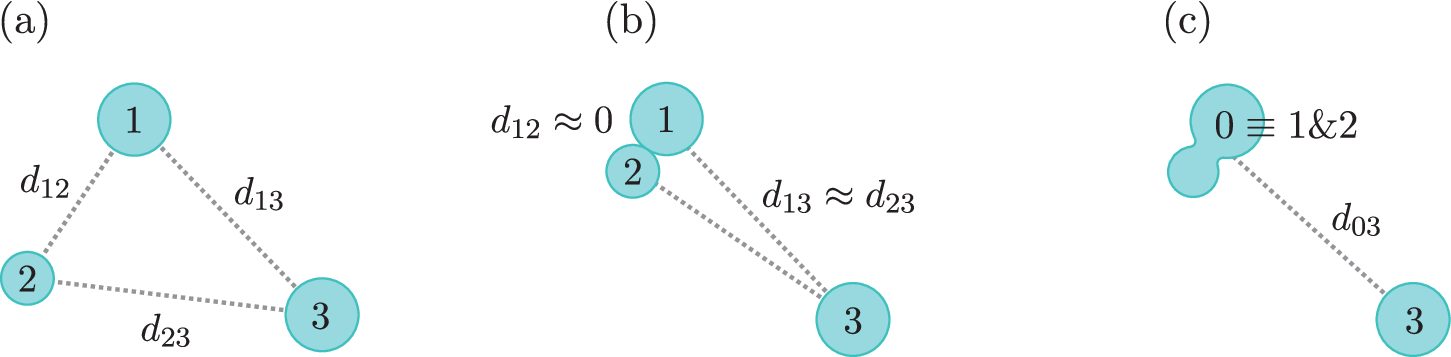}
\caption{{(a)}, a general three-category setup.
{(b)}, a three-category setup when Category 1 and Category 2 can be regarded as virtually identical.
{(c)}, a two-category setup whose physical contents are the same as diagram {(b)}.
The principle of splitting/merging invariance (Condition \ref{cond:split/merg}) requires that the diversity indices computed from diagram {(b)} and diagram {(c)} agree.}
\label{fig:split-merge-inv}
\end{figure}

\noindent
We briefly comment particularly on Condition \ref{cond:split/merg}, which we refer to as the principle of invariance under splitting or merging of identical categories, or the principle of \q{splitting/merging invariance} for short.
Fig.\ \ref{fig:split-merge-inv} illustrates how this principle works for the simplest \q{three-category $\leftrightarrow$ two-category} case.
Fig.\ \ref{fig:split-merge-inv}{(a)} represents a general three-category setup with diversity $\iD(\{p_{1},\,p_{2},\,p_3\},\{d_{12},\,d_{13},\,d_{23}\})$.\footnote{Here and below we omit to display terms like $d_{ij}$ with $i>j$ as well as $d_{ii}=0$ to save space.}
Fig.\ \ref{fig:split-merge-inv}{(b)} represents the same three-category setup but now in the limit when the dissimilarity between Category 1 and Category 2 is negligible or undetectable, for which the diversity function reduces to $\iD^{\mathrm{i}}=\iD(\{p_{1},\,p_{2},\,p_3\},\{d_{12}=0,\,d_{13}=d_{23}\})$.
By contrast, Fig.\ \ref{fig:split-merge-inv}{(c)} represents a two-category setup with diversity $\iD^{\mathrm{ii}}=\iD(\{p_{0},\,{p_3}\},\{d_{03}\})$, whose physical contents---before labelling and putting into mathematical terms---can be regarded as the same as those of Fig.\ \ref{fig:split-merge-inv}{(b)}.
In this particular case, the principle of splitting/merging invariance (Condition \ref{cond:split/merg}) states that $\iD^{\mathrm{i}}=\iD^{\mathrm{ii}}$ holds under identification of $p_{1}+p_{2}=p_{0}$ and $d_{13}=d_{23}=d_{03}$.
Although many of the diversity indices considered in earlier works do not generally satisfy this invariance property, we postulate it as an inherent symmetry of diversity since any diversity indices without this property fail to properly reflect the aspect of between-category dissimilarity.
It is noteworthy that the diversity index derived heuristically in Ref.\ \cite{Leinster12} satisfies this principle of splitting/merging invariance.
Our axiomatic construction of the diversity index uses this invariance property to constrain and refine the functional form of $\iD$, thereby yielding a larger class of diversity functions than considered in \cite{Leinster12}; see below.\\

Let us consider a general assemblage with distribution $\{p_{i}\}$ and the set of between-category dissimilarities $\{d_{ij}\}$, where $i,\,j\in \cS$.
Here we do not presume $P=\sum_{i\in \cS}p_{i}$ to be unity so that the subsequent discussions and the resulting formulae equally hold when $\cS$ represents only part of those categories available.
Corresponding to each such observed state, a unique normalised state---a hypothetical assemblage containing $\iD$ equiprevalent categories that are maximally dissimilar to each other---can be characterised by the uniform distribution $p(t)=P/\iD$ with $\int_{\cS}p(t)dt=1$ and $\int_{\cS}dt=\iD$ as before.
The condition that $\iD$ represents the effective diversity this time is then given by the following equivalence relation:
\begin{equation}\label{iD_def_elementary}
M_{q-1}\kko{\big\{A_{i}(\{p_{i}\},\{d_{ij}\})\big\} \Big| \Big\{\f{p_{i}}{P}\Big\}}
= \cM_{q-1}\kko{\hf{1}{\iD} \Big| \hf{1}{\iD}}\,,
\end{equation}
where $A_{i}$ is a continuous function of both the category abundances $\{p_{i}\}$ and the between-category dissimilarities $\{d_{ij}\}$.
By solving this equation, we obtain
\begin{equation}\label{iD_def_elementary2}
\iD(\{p_{i}\},\{d_{ij}\})=\Bigg[\sum_{i\in\cS}\f{p_{i}}{P}\,A_{i}(\{p_{i}\},\{d_{ij}\})^{q-1}\Bigg]^{\f{1}{1-q}}\,.
\end{equation}
To refine the functional form of $\iD$, we make the following ansatz for $A_{i}$:
\begin{equation}\label{A-W}
A_{i}(\{p_{i}\},\{d_{ij}\})=\sum_{j\in \cS} W(p_{j}|d_{ij})\,,
\end{equation}
where $W$ is again a continuous function of both $\{p_{i}\}$ and $\{d_{ij}\}$, and feed it into (\ref{iD_def_elementary2}).
A small calculation shows that a necessary condition for $\iD^{\mathrm{i}}=\iD^{\mathrm{ii}}$ (Condition \ref{cond:split/merg}) in terms of $W$ is given by
\begin{equation}\label{cond:W}
\sum_{j=1,\,2}W(p_{j}|d_{0i})=W(p_{0}|d_{0i})
\end{equation}
for all $i\in\cS$.
Assuming factorisation $W(p_{i}|d_{ij})=g(p_{i})h(d_{ij})$ with $g$ and $h$ continuous functions of the respective arguments, the condition (\ref{cond:W}) implies $g(p_{1})+g(p_{2})=g(p_{0})$, whose solution is given by $g(p)=cp$ with $c$ (here non-zero) constant.
Subsequently, Condition \ref{cond:reduceHill}, stating that $A_{i}$ reduces to $p_{i}/P$ if $d_{ij}=1-\delta_{ij}$, requires that $cp_{i}h(0)+c\sum_{j\,(\neq i)} p_{j}h(1)= p_{i}/P$ holds for each $i$, which implies $cPh(0)=1$ and $h(1)=0$.
By defining $K(x)\coloneqq cPh(x)$, these conditions are rewritten as, respectively,
\begin{equation}\label{K(0)K(1)}
K(1)=0 \quad\text{and}\quad K(0)=1\,.
\end{equation}
In terms of this function $K(\cdot)$, the dissimilarity-based extended Hill numbers family is finally written as
\begin{align}\label{D-elementary}
{}^{q}\!\iD(\{p_{i}\},\{d_{ij}\})
=\begin{cases}
\, \ds \kko{\sum_{i\in \cS} \f{p_{i}}{P}\Bigg(\sum_{j\in \cS} \f{p_{j}}{P}K(d_{ij})\Bigg)^{q-1}}^{\f{1}{1-q}} & (q\neq 1) \\[4mm]
\, \ds \exp\Bigg[-\sum_{i\in\cS}\f{p_{i}}{P}\ln\Bigg( \sum_{j\in\cS}\f{p_{j}}{P}K(d_{ij}) \Bigg)\Bigg]  & (q=1)
\end{cases}\,.
\end{align}
Here, Condition \ref{cond:even_p} requires $K(d_{ij})$ to be a continuous monotonically decreasing function of $d_{ij}$ satisfying the conditions (\ref{K(0)K(1)}); this feature makes it appropriate to refer to $K(d_{ij})$ as the \emph{effective affinity} between categories $i$ and $j$, and therefore, hereafter we refer to (\ref{D-elementary}) as the affinity-based extended diversity index.
It can also be checked that Conditions \ref{cond:max} and \ref{cond:min} are readily satisfied.
Subsequently, the affinity-based extended Tsallis entropy is obtained as
\begin{align}\label{Tsallis_d}
\hspace{-1mm}\cH_{q}(\{p_{i}\},\{d_{ij}\})
&=\ln_{q}{}^{q}\!\iD(\{p_{i}\},\{d_{ij}\})\no\\
&=\begin{cases}
\, \ds \f{1}{1-q}\sum_{i\in\cS}\f{p_{i}}{P}\Bigg[\Bigg(\sum_{j\in\cS}\f{p_{j}}{P}K(d_{ij})\Bigg)^{q-1}-1\Bigg] & (q\neq 1) \\[4mm]
\, \ds -\sum_{i\in\cS}\f{p_{i}}{P}\ln\Bigg( \sum_{j\in\cS}\f{p_{j}}{P}K(d_{ij}) \Bigg)  & (q=1)
\end{cases}
\end{align}
with $\cH_{1}$ the affinity-based extension of Boltzmann--Gibbs entropy (\ref{Shannon}).

The diversity index (\ref{D-elementary}) can be regarded as a unique extension of the Hill numbers family enhanced by the newly introduced function $K(\cdot)$, which we refer to as the \emph{kernel function}.
Note that it is this kernel structure that introduces the endogenous dependence of the dissimilarity factor on the quantification of diversity/entropy.
If the value of $K(d_{ij})$ is identified as an exogenous \q{similarity} variable, $Z_{ij}\in [0,1]$, defined such that $Z_{ii}=1$ for all $i$, then the diversity index (\ref{D-elementary}) reduces to that considered in \cite{Leinster12,Leinster16}, which has no kernel structure (see also Fig.\ \ref{fig:comparison} of \ref{app:diversity indices}).
In the following subsection, we will see that the kernel function plays a crucial role in establishing the affinity-based diversity formula that is not only applicable to the \q{elementary} level analysis as thus far but also to any aggregate meta-level analyses, where the building blocks of diversity are coarse-grained (\q{effective}) quantities represented by $\{P_{I}\}$ and $\{\bar{d}_{IJ}\}$ with each coarse-grained category ($I,\,J,\,\dots$) consisting of the elementary-level categories ($i,\,j,\,\dots$).
Recall that we had the relation $d_{ii}=0$, by definition, at the elementary level.
It will be shown that the same form of relation at the aggregate level, \q{${d}_{II}=0$}, no longer holds in general.
This means that the framework of \cite{Leinster12,Leinster16}, which is based on the definition that $Z_{II}=1$, implying ${d}_{II}=K^{-1}(Z_{II})=0$, needs to be modified and enhanced for it to be applicable to an aggregate level diversity/entropy analysis.

\begin{figure}
\centering
\includegraphics[width=16.3cm,clip]{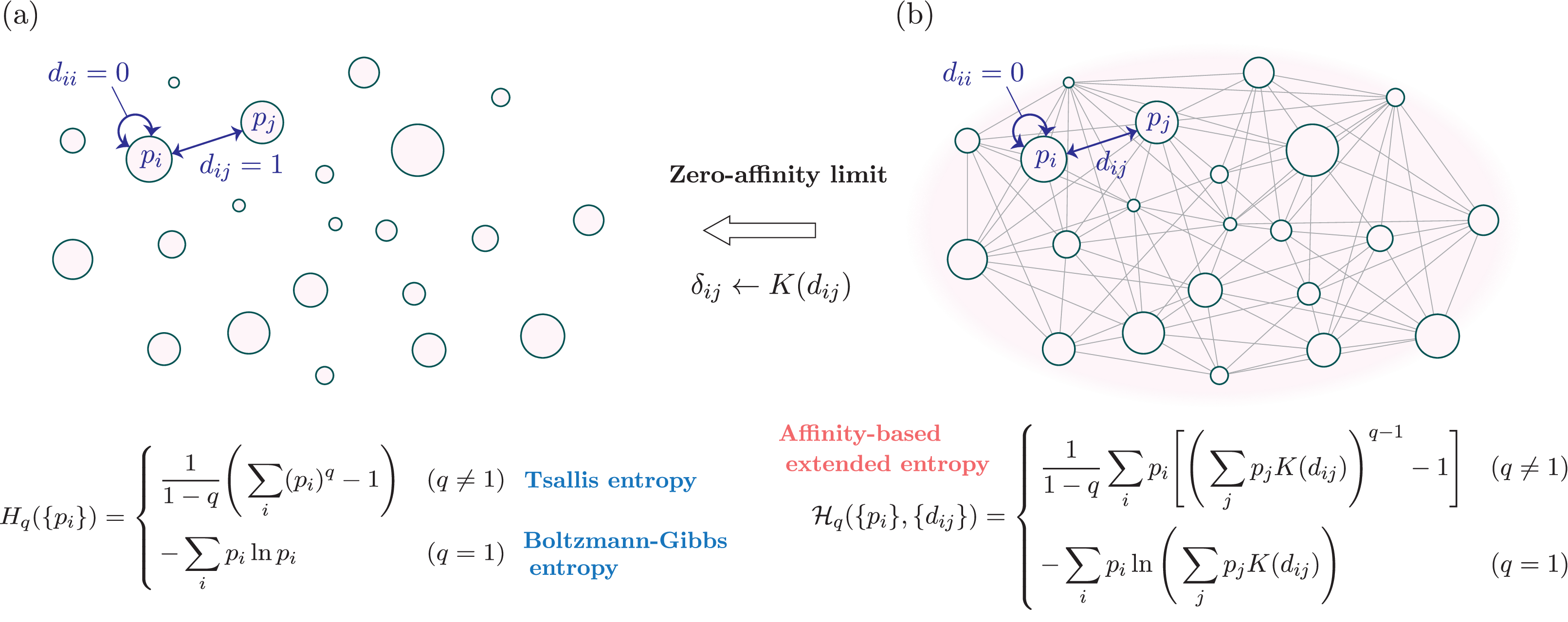}
\caption{(a), a statistical mechanical system characterised only by the probability distribution, $\{p_{i}\}$, without any between-state affinities, which is described by Tsallis and Boltzmann--Gibbs entropies.
(b), another statistical mechanical system characterised not only by the probability distribution but also by the between-state dissimilarities, $\{d_{ij}\}$, which is described by the affinity-based extended entropy.
Picture (a) is obtained as the zero-affinity limit, $K(d_{ij})\to \delta_{ij}$, of Picture (b).}
\label{fig:reduction}
\end{figure}

A few more remarks concerning the above results are in order.
First, up to this point, the kernel function can be of an arbitrary functional form as long as it is compatible with the properties derived from Conditions \ref{cond:conti}--\ref{cond:min}, including the conditions (\ref{K(0)K(1)}).
Its functional form will be shown to be constrained by the special property called the Nesting Principle introduced shortly.
Second, there are two routes for the diversity index (\ref{D-elementary}) and the extended Tsallis entropy (\ref{Tsallis_d}), respectively, to reduce to the Hill numbers family (\ref{Hill_num}) and the standard Tsallis and Boltzmann--Gibbs entropies, (\ref{Tsallis}) and (\ref{Shannon}).
These are either due to the exogenous reduction, $d_{ij}\to 1-\delta_{ij}$, or due to the endogenous reduction by way, for instance, of the limit $r\to 0$ of the kernel function given by $K_{r}(d_{ij})=1-(d_{ij})^{r}$ (keeping $\{d_{ij}\}$ general).
In either case, it follows that $K(d_{ij})\to \delta_{ij}$; see Fig.\ \ref{fig:reduction} for a schematic representation ($P=1$ is set for simplicity).
Third, note that the extended entropy (\ref{Tsallis_d}) is not generally a concave function of probabilities $\{p_{i}\}$.
However, this is not an issue as long as the diversity index (\ref{D-elementary}) satisfies Conditions \ref{cond:conti}--\ref{cond:min} along with other prerequisites discussed above.
Fourth, a more refined axiomatisation than presented in this work could have shed better light on the mathematical structure embedded in the effective diversity/entropy.
In the zero-affinity limit, the Shannon--Khinchin axioms \cite{Shannon64,Khinchin13} in information theory and the Shore--Johnson axioms \cite{ShoreJohnson80,ShoreJohnson81} in statistical inference theory are well-established axiomatisation for the Boltzmann--Gibbs--Shannon ($q=1$) case, and their non-extensive ($q\neq 1$) generalisation \cite{Santos97,Abe00,Suyari04,Furuichi05} is also known.
These axiomatisations of diversity/entropy in the zero-affinity case have been, in effect, encapsulated in our Condition \ref{cond:reduceHill}.
We expect the current affinity-based extended diversity/entropy, along with other types of extended entropies (see \ref{app:divergence}), to be also derivable from as \q{thrifty} and structured a set of axioms as possible.

\subsection{The Nesting Principle\label{sec:NP}}

So far, we have considered an assemblage (\q{a system}) of categories (\q{microstates}) defined at the most fundamental level, in which each category was homogeneous in the sense that the constituent elements within each category were all identical.
Now, suppose that the assemblage is partitioned into sub-assemblages, each composed of the fundamental categories.
Then a natural question to ask is:
Is it possible to regard these sub-assemblages as \q{coarse-grained categories} and use them as new building blocks to (re-)evaluate the effective diversity?
If it is possible, then under what condition(s) the resulting diversity index is consistent with that computed at the elementary level?
Put differently, are any further constraints on the functional form of the general affinity-based diversity (\ref{D-elementary}) required for the \q{grouping-invariance}, or the \q{nesting-invariance} property to hold?
Below we provide the answers to these questions by deriving a diversity formula that is applicable at any scale of the granularity of entities, from any different level or layer of categorisation, and with any arbitrary partitioning or aggregation.\\

Suppose that an assemblage is partitioned into $N$ sub-assemblages, or \q{groups}, labelled by $I\in\cN\coloneqq\{1,\,\dots,\,N\}$, each composed of $n_{I}$ categories.
The sum of relative abundances of categories within each group is given by 
\begin{equation}\label{def:P}
P_{I}\coloneqq \sum_{i\in\cS_{I}}p_{i}\,, 
\end{equation}
where $\cS_{I}\coloneqq\{1,\,\dots,\,n_{I}\}$.
We denote the effective dissimilarity between groups $I$ and $J$ as $\bar{d}_{IJ}$, whose explicit form in terms of the elementary-level variables $\{p_{i}\}$ and $\{d_{ij}\}$ is yet to be determined.
Still, $\{P_{I}\}$ and $\{\bar{d}_{IJ}\}$ denote the coarse-grained equivalents of the elementary-level $\{p_{i}\}$ and $\{d_{ij}\}$, respectively.
There is another effective quantity defined at the coarse-grained level: it is the effective diversity of each group.
Here, the original homogeneous categories labelled by $\{i\}$, defined at the elementary level with each having $\iD_{i}=1$ (by definition), are integrated and averaged within each group $I$ to give the \q{partial} effective diversity, $\iD_{I}\coloneqq \iD(\{p_{i}\},\{d_{ij}\})_{i,\,j\in\cS_{I}}$.
It therefore holds that $1\leq \iD_{I}\leq n_{I}$ for each $I\in\cN$ by construction.

\begin{figure}
\centering
\includegraphics[width=14.7cm,clip]{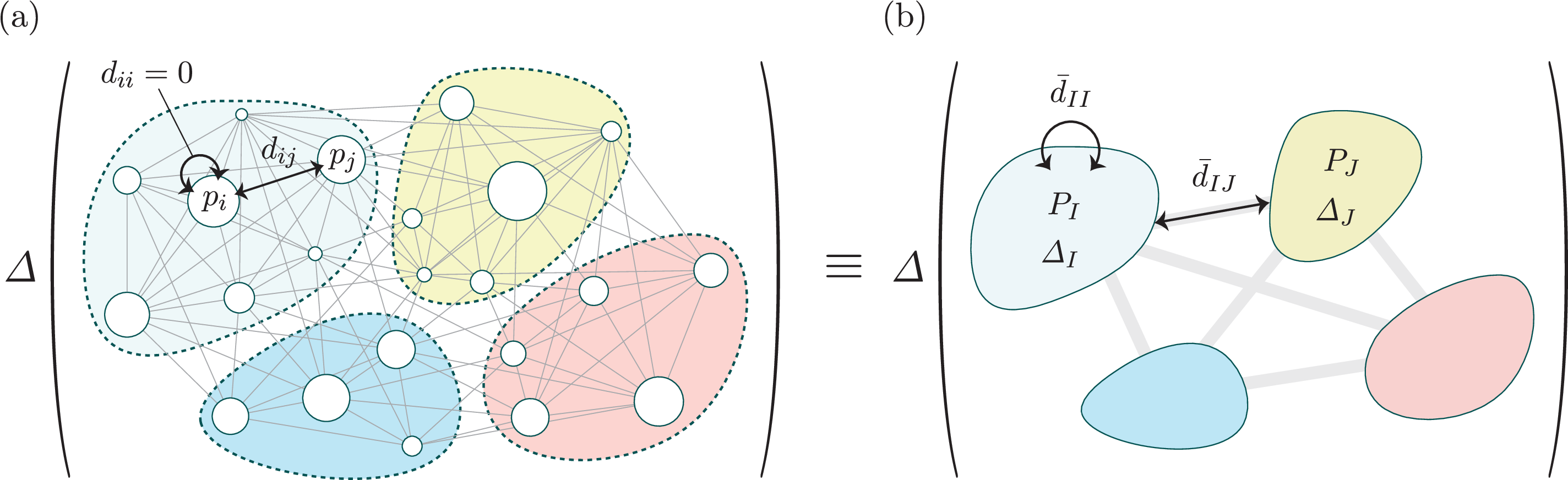}
\caption{A schematic representation of the Nesting Principle.
(a), the elementary level picture described by $\{p_{i}\}$ and $\{d_{ij}\}$.
(b), an aggregate meta-level picture obtained by \q{coarse-graining} the elementary level picture, described by $\{P_{I}\}$, $\{\bar{d}_{IJ}\}$ (and $\{\iD_{I}\}$).
The Nesting Principle states that the diversity indices computed from the two pictures (a) and (b) agree.}
\label{fig:nesting}
\end{figure}

Now, we introduce a new principle called the \emph{Nesting Principle}, which strengthens the principle of splitting/merging invariance introduced earlier.
Specifically, it states that the effective diversity computed from the original elementary variables (see Eq.\ (\ref{D-elementary})),\footnote{Hereafter, the expressions in the limit $q\to 1$ is not always displayed, for ease of reading.
However, the limiting case readily follows by the continuity of power means $M_{q-1}$ with respect to the order parameter $q$.}
\begin{align}\label{aggregateD}
\iD_\mathrm{elem} (\{p_{i}\},\{d_{ij}\})=\Bigg[ \sum_{\mathrm{all}\, i}\f{p_{i}}{P}\Bigg( \sum_{\mathrm{all}\, j}\f{p_{j}}{P} K(d_{ij})\Bigg)^{q-1} \Bigg]^{\f{1}{1-q}}\,,
\end{align}
agrees with that computed from the \q{coarse-grained} effective variables defined at the aggregate meta-level,
\begin{align}\label{coarse-grainedD}
\iD_\mathrm{nest} (\{P_{I}\},\{\bar{d}_{IJ}\})
=\Bigg[ \sum_{I\in\cN}\f{P_{I}}{P}\Bigg( \sum_{J\in\cN}\f{P_{J}}{P} K(\bar{d}_{IJ})\Bigg)^{q-1} \Bigg]^{\f{1}{1-q}}\,,
\end{align}
where $\sum_{\mathrm{all}\, i}=\sum_{I\in\cN}\sum_{i\in\cS_{I}}$ and $P=\sum_{I\in\cN}P_{I}=\sum_{\mathrm{all}\, i}p_{i}$ as before.
This means that the diversity formula (\ref{D-elementary}) is universally applicable to any level of nesting structures, both for the elementary-level building blocks, $\{p_{i}\}$ and $\{d_{ij}\}$, and for the \q{coarse-grained} quantities, $\{P_{I}\}$ and $\{\bar{d}_{IJ}\}$, according to each nested level.
See Fig.\ \ref{fig:nesting} for a schematic representation.
This equivalence relation, \q{$\iD_\mathrm{elem}=\iD_\mathrm{nest}$}, is shown to strongly constrain the functional form of the affinity-based diversity index (\ref{D-elementary}) and hence of the associated entropy function.
Before showing this, we note that the reciprocal of Eq.\ (\ref{D-elementary}) can be written as
\begin{equation}\label{D-elementary2}
\f{1}{\iD_{I}}=
M_{q-1}\Big[\Big\{\sum\nolimits_{j\in\cS_{I}}\f{p_{j}}{P_{I}}K(d_{ij})\Big\} \Big| \Big\{\f{p_{i}}{P_{I}}\Big\}\Big]\,.
\end{equation}
Here, first the standard arithmetic mean of $K(d_{ij})$ with nominal weight $p_{j}/P_{I}$ is taken over all $j$ in group $I$, and subsequently, a $(q-1)$-th power mean is taken over all $i$ in the same group $I$.
This quantity has an interpretation as a certain generalised correlation sum defined for group $I$.
Therefore, the corresponding cross-correlation sum is considered between two distinct groups $I$ and $J$, which is adopted as the definition of $K(\bar{d}_{IJ})$, i.e.\
\begin{equation}\label{K(d)_IJ}
K(\bar{d}_{IJ})=
M_{q-1}\Big[\Big\{\sum\nolimits_{j\in\cS_{J}}\f{p_{j}}{P_{J}}K(d_{ij})\Big\} \Big| \Big\{\f{p_{i}}{P_{I}}\Big\}\Big]\,.
\end{equation}
Namely, it is obtained by first taking an arithmetic mean over all $j$ in one of the groups, $J$, then taking a $(q-1)$-th power mean over all $i$ in the other group, $I$.
In effect, this averaged quantity can be viewed as representing the effective affinity between two groups $I$ and $J$.
Note that the reciprocal relation $K(\bar{d}_{IJ})=K(\bar{d}_{JI})$ only holds either when $q=2$ in the light of the definition (\ref{K(d)_IJ}), or when each group contains only a single category, i.e.\ $n_{I}=\iD_{I}=1$ and $n_{J}=\iD_{J}=1$.
By setting $I=J$ in (\ref{K(d)_IJ}) and comparing the resulting expression with (\ref{D-elementary2}), the following consistency condition is obtained:
\begin{equation}\label{d_II}
K(\bar{d}_{II})=\f{1}{\iD_{I}}\,.
\end{equation}
This relation is simple yet insightful:
The effective diversity of a system is given by the inverse self-correlation sum defined for the same system.
In the light of the condition (\ref{K(0)K(1)}), it also indicates that $\bar{d}_{II}=0$ is satisfied only if the constituents within $I$ are homogeneous, i.e.\ provided that $\iD_{I}=1$.
This is a contrasting situation to what was defined at the elementary level, i.e.\ $d_{ii}=0$ for all $i$.

We now show how the Nesting Principle constrains the functional form of the diversity function (\ref{D-elementary}).
It suffices to show for the $N=2$ case.
Let us consider two groups $I$ and $J$, and define
\begin{align}
\cU_{i}\coloneqq \f{p_{i}}{P}\Bigg(\sum_{j\in\cS_{I}}\f{p_{j}}{P}K(d_{ij})\Bigg)^{1/s}\quad 
\text{and}\quad 
\cV_{i}\coloneqq \f{p_{i}}{P}\Bigg(\sum_{j\in\cS_{J}}\f{p_{j}}{P}K(d_{ij})\Bigg)^{1/s}
\end{align}
with $s\coloneqq 1/(q-1)$.
Then, on the one hand, the diversity index at the elementary level, (\ref{aggregateD}), satisfies
\begin{equation}\label{delta_a}
\big(\iD_\mathrm{elem}\big)^{-1/s}=\sum_{i\in\cS_{I}}\Big(\cU_{i}{}^{s}+\cV_{i}{}^{s}\Big)^{1/s}+\sum_{j\in\cS_{J}}\Big(\cU_{j}{}^{s}+\cV_{j}{}^{s}\Big)^{1/s}\,,
\end{equation}
while on the other hand, the diversity index at the aggregate meta-level, (\ref{coarse-grainedD}), satisfies
\begin{align}\label{delta_b}
\big(\iD_\mathrm{nest}\big)^{-1/s}=\Bigg[\Bigg(\sum_{i\in\cS_{I}}\cU_{i}\Bigg)^{s}+\Bigg(\sum_{i\in\cS_{I}}\cV_{i}\Bigg)^{s}\,\Bigg]^{1/s}
+\Bigg[\Bigg(\sum_{j\in\cS_{J}}\cU_{j}\Bigg)^{s}+\Bigg(\sum_{j\in\cS_{J}}\cV_{j}\Bigg)^{s}\,\Bigg]^{1/s}\,.
\end{align}
Using Jensen's inequality for a two-variable function $f(x,y)=(x^{s}+y^{s})^{1/s}$, it follows that\footnote{The equality $\iD_\mathrm{elem}/\iD_\mathrm{nest}=1$ also holds if $\cU_{i}$ and $\cV_{i}$ are constant across $i\in\cS_{I}$, and $\cU_{j}$ and $\cV_{j}$ are constant across $j\in\cS_{J}$. However, here we exclude this particular case and focus on the range of $s$ (or $q$) with an arbitrary set of $\cU$s and $\cV$s.}
\begin{equation*}
\f{\iD_\mathrm{elem}}{\iD_\mathrm{nest}}
\left\{\begin{aligned}
&>\,\\[-2mm]
&=\,\\[-2mm]
&<\,
\end{aligned}\right\}
1
\quad\text{if}\quad
\left\{\begin{aligned}
&~0<s<1\,,\\[-2mm]
&~s=1\,,\\[-2mm]
&~s>1,~s\leq -1\,.
\end{aligned}\right.
\end{equation*}
Consequently, the Nesting Principle is satisfied if and only if $s=1$, or $q=2$; otherwise the diversity $\iD_\mathrm{nest}$ computed at the aggregate meta-level either overestimates ($q\in [0,2)$) or underestimates ($q\in (2,\infty)$) the \q{true} diversity $\iD_\mathrm{elem}$ computed at the elementary level.
Thus, the Nesting Principle requires the diversity function to take a quadratic form in relative abundances of categories, in which case the formula for the effective affinity between two groups $I$ and $J$ is given by, from Eq.\ (\ref{K(d)_IJ}),
\begin{equation}\label{formula2}
K(\bar{d}_{IJ})=\sum_{i\in\cS_{I}}\sum_{j\in\cS_{J}}\f{p_{i}}{P_{I}}\f{p_{j}}{P_{J}}K(d_{ij})\,.
\end{equation}
As the last step of our construction of the general diversity formula, we adopt the following definition of the effective dissimilarity:
\begin{align}\label{def:d_IJ}
\bar{d}_{IJ}(\{p_{i}\},\{d_{ij}\})
=M_{r}\Bigg[\{d_{ij}\} \Bigg| \Bigg\{\f{p_{i}}{P_{I}}\f{p_{j}}{P_{J}}\Bigg\}\Bigg]_{(i,j)\in\cS_{I}\times\cS_{J}}
=\Bigg[\sum_{i\in\cS_{I}}\sum_{j\in\cS_{J}}\f{p_{i}}{P_{I}}\f{p_{j}}{P_{J}}\,\big(d_{ij}\big)^{r}\Bigg]^{1/r}\,.
\end{align}
Here the parameter $r$ is introduced to control the effective dissimilarity measure's sensitivity to the elementary between-category dissimilarities, to which little attention has been paid before the present work.
Notice that the key relation (\ref{d_II}) implies that not only $\iD_{I}$ but also $\bar{d}_{II}$ is invariant under the normalisation procedure (the reader may recall how the normalisation procedure worked in generating effective quantities; see Section \ref{subsec:preliminaries}).
By setting $I=J$ in (\ref{def:d_IJ}) and using the mapping from the observed state to the corresponding normalised state as $p_{i}/P_{I}\mapsto 1/\iD_{I}$, $(d_{ij})^{r}\mapsto 1-\delta(t'-t)$ with $\delta(\cdot)$ the Dirac delta function, along with $\sum_{i\in\cS_{I}}\sum_{j\in\cS_{I}}\mapsto \iint_{\cS\times\cS} dtdt'$, it can be deduced that
\begin{align}\label{d_II_int}
(\bar{d}_{II})^{r}
=\iint_{\cS\times\cS}\f{dt}{\iD_{I}}\f{dt'}{\iD_{I}}\big(1-\delta(t'-t)\big)
=1-\f{1}{\iD_{I}}\,.
\end{align}
Subsequently, this relation and the previous equivalence relation (\ref{d_II}) are solved to give 
\begin{equation}\label{def:K(x)}
K_{r}(x)=1-x^{r}\,.
\end{equation}
Here the kernel function is indexed by the subscript $r$ to make its dependence on the scaling parameter explicit.
One can check that (\ref{def:K(x)}) solves the functional equation (\ref{formula2}) with (\ref{def:d_IJ}) for general sets of $\{p_{i}\}$ and $\{d_{ij}\}$.
Plugging (\ref{def:K(x)}) and $q=2$ in (\ref{coarse-grainedD}), we finally arrive at the following affinity-based, nesting-invariant diversity formula:
\begin{equation}\label{coarse-grainedD2}
\iD_{\mathrm{nest}}^{(r)}(\{P_{I}\},\{\bar{d}_{IJ}\})
=\Bigg[1-\sum_{I,\,J\in\cN}\f{P_{I}}{P}\f{P_{J}}{P}\big(\bar{d}_{IJ}\big)^{r}\Bigg]^{-1}\,,
\end{equation}
where the effective (coarse-grained) quantities $\{P_{I}\}$ and $\{\bar{d}_{IJ}\}$ are defined as in (\ref{def:P}) and (\ref{def:d_IJ}), respectively.
When each group is homogeneous in category composition (i.e.\ $n_{I}=\iD_{I}=1$ for all $I\in\cN$), the above formula reduces to the elementary-level formula,
\begin{equation}\label{coarse-grainedD2-e}
\iD_{\mathrm{elem}}^{(r)}(\{p_{i}\},\{d_{ij}\})
=\Bigg[1-\sum_{i,\,j\in\cN}\f{p_{i}}{P}\f{p_{j}}{P}\big(d_{ij}\big)^{r}\Bigg]^{-1}\,,
\end{equation}
and the associated extended Tsallis entropy compatible with the Nesting Principle is given by\footnote{Interestingly, it is essentially the same quantity as considered in Ref.\ \cite{Rao82} in the context of population biology.}
\begin{align}\label{Tsallis_d_nest}
\cH_{q=2}(\{p_{i}\},\{d_{ij}\})
=\ln_{q=2}\iD_{\mathrm{elem}}^{(r)}
=\sum_{i,\,j\in\cN}\f{p_{i}}{P}\f{p_{j}}{P}(d_{ij})^{r}\,.
\end{align}

The diversity formula (\ref{coarse-grainedD2-e}) can be seen to extend the diversity measure considered in \cite{Leinster12,Leinster16} in two ways.
One is by making it universally applicable to any level of nesting structure, as discussed, and the other is by adding a new degree of freedom to control the weight given to between-category dissimilarity.
The latter is achieved by the scaling parameter $r$ that determines the measures' emphasis on similar or dissimilar categories, generating a new one-parameter continuous spectrum of diversity/entropy.
In the limit $r\to 0$, the diversity measure gives as much importance to almost identical category pairs as maximally dissimilar ones, which corresponds to the Hill numbers case (i.e.\ $\lim_{r\to 0}K_{r}(d_{ij})=\delta_{ij}$).
If $r\in(0,1)$, then the diversity measure assigns a greater weight to relatively similar category pairs, whereas if $r\in(1,\infty)$, then it assigns a greater weight to relatively dissimilar category pairs, with the limiting case $r\to\infty$ reflecting only the prevalence of the most dissimilar category pairs.
By contrast, the order parameter $q$, which indexes the Hill numbers family (\ref{Hill_num}) to control the measures' emphasis on rare or common categories, has to be fixed to $q=2$ for the Nesting Principle to generally hold; see Fig.\ \ref{fig:comparison} of \ref{app:diversity indices}.
All these points answer the question of how to formulate the diversity index that consistently incorporates all the essential aspects of diversity, including both the exogenous aspects ($\{p_{i}\}$ and $\{d_{ij}\}$) and the endogenous aspects ($q$ and $r)$.\footnote{There is a caveat regarding the range of the parameter $r$ and its implication on the prerequisite for dissimilarity variables.
Recall that the dissimilarity variables $\{d_{ij}\}$ are assumed to obey the triangle inequality; see the argument below Eq.\ (\ref{general_iD}).
For $r \in [1,\infty)$, it immediately follows from the definition (\ref{def:d_IJ}) of the effective dissimilarities $\{\bar{d}_{IJ}\}$ and Minkowski's inequality that the same triangle inequality also holds at the aggregate level: $\bar{d}_{IJ}+\bar{d}_{JK}\geq \bar{d}_{IK}$ for arbitrary triplets of coarse-grained categories  $(I,\,J,\,K)$.
For $r\in (0,1)$, however, the usual triangle inequality is not necessarily satisfied.
In this region of $r$, the effective dissimilarities are shown to obey a certain relaxed version of the triangle inequality.
The author would like to thank R.\ Suzuki for discussions on this point.}

It is worth noting that, for the nesting-invariant $q=2$ case, the computation of the diversity index at a given nested level can be conducted without knowing about the \q{internal structures} of the building blocks defined at the nested level.
Only needed are the information of the effective quantities $\{P\}$ and $\{\bar{d}\}$ (including $\{\iD\}$) at the nested level one is at, and hence the microscopic origin of these coarse-grained quantities (e.g.\ how point probabilities are distributed at the finer-grained level, to what degree the microstates are correlated with each other, or how micro-structures are integrated into the current macro-structure) are not necessary to calculate the universal diversity index.\footnote{All these features resemble those seen in the theory of renormalisation group, which plays a vital role in some areas of physics.
For instance, in the statistical physics of phase transitions, the renormalisation group is defined based on recursive averaging over short distance degrees of freedom.
In the current context, the \q{depth} of the nesting can be seen to correspond to the renormalisation scale.
Specifically, the composite $P^{2}/\iD$ can be seen to play the role of the running coupling in Gell-Mann and Low type renormalisation group equations.
The coarse-graining procedure by means of the weighted correlation sum (\ref{def:d_IJ}) can be viewed as defining a certain beta function: $\{\{P_{I'}\},\{\bar{\bar{d}}_{I'J'}\}\}=\beta(\{\{P_{I}\},\{\bar{d}_{IJ}\}\})$, which induces the renormalisation flow.}

In closing this section, we emphasise that the condition $q=2$ for the nesting-invariance is relaxed in the zero-affinity limit.
In this case, the diversity index defined at the elementary level, (\ref{aggregateD}), and that defined at the aggregate meta-level, (\ref{coarse-grainedD}), are related as (recalling the previous notation to use $D$ instead of $\iD$ to represent Hill numbers):
\begin{align}
\big(D_\mathrm{elem}\big)^{1-q}
&\eq\sum_{\mathrm{all}\, i}\bigg(\f{p_{i}}{P}\bigg)^{q}
=\sum_{I\in\cN}\sum_{i\in\cS_{I}}\bigg(\f{p_{i}}{P_{I}}\bigg)^{q} \bigg(\f{P_{I}}{P}\bigg)^{q}
=\sum_{I\in\cN}(D_{I})^{1-q}\bigg(\f{P_{I}}{P}\bigg)^{q}\no\\
&=\sum_{I\in\cN}\big(K(\bar{d}_{II})\big)^{q-1}\bigg(\f{P_{I}}{P}\bigg)^{q}
=\sum_{I\in\cN}\f{P_{I}}{P}\bigg(\f{P_{I}}{P}K(\bar{d}_{II})\bigg)^{q-1}
\eq\big(D_\mathrm{nest}\big)^{1-q}\,,
\end{align}
i.e.\ $D_\mathrm{elem}=D_\mathrm{nest}$, ensuring that the Nesting Principle holds for arbitrary $q$.
This means that the Nesting Principle is always compatible with Tsallis and Boltzmann--Gibbs statistical mechanics, in which no degrees of freedom associated with between-state affinity are considered.
Another remark concerns the threshold of detectable dissimilarity or affinity, which inherently limits the validity of the diversity/entropy formula.
In the above theoretical framework, it is implicitly assumed that the dissimilarity or affinity can be measured with infinite precision.
In practice, however, such precision is not available for various technical or intrinsic reasons.
Consequently, in reality, the Nesting Principle will only hold approximately even with $q=2$ or with $K(d_{ij})=\delta_{ij}$, albeit it could be very close to being exact.

\section{Applications to Non-Extensive Statistical Mechanics\label{subsec:non-extensive}}

Tsallis statistical mechanics \cite{Tsallis88,Tsallis94,Tsallis99,Tsallis09,Abe-Okamoto01} has been reported to be successful in various situations in which Boltzmann--Gibbs statistical mechanics is no longer valid, e.g.\ when with long-range interactions, long-term microscopic memory (non-Markovian processes), and multifractal relevant spacetime or phase space structures; see \cite{TsallisBibList} for an extensive list of references.
In this section, we investigate the implications of the affinity-based extended Tsallis entropy for the non-extensive statistical mechanical systems at equilibrium.
In particular, the equilibrium probability distributions are shown to be no longer given in the Tsallis' $q$-deformed forms but in a form modified by the presence of between-state affinities, whose physical origin will depend on the application context.
The present application to statistical mechanics is largely motivated by theoretical (mathematical-physical) considerations.
However, it would be interesting if future research could verify some of the consequences of the theoretical model described below.
In fact, as will be shown shortly, the extended theory predicts that the behaviours of some measurable thermodynamic quantities at equilibrium are affected by between-state affinities.
If such a deviation from the standard zero-affinity theory could be observed experimentally, it could imply that the between-state affinity actually exists in the system.

Hereafter in this section we mainly use the \q{distance} variable $\rho_{ij}\in [0,\infty)$, which is obtained from the dissimilarity variable $d_{ij}$ via the transformation $(d_{ij})^{r}=1-e^{-(\rho_{ij})^{\eta}}$ with $\eta\in (0,\infty)$.
The effective affinity in the formula (\ref{D-elementary}) is then given in terms of the distance variable as
\begin{equation}\label{d<->r2}
K(d_{ij})=\kappa(\rho_{ij})= e^{-(\rho_{ij})^{\eta}}\,,
\end{equation}
so that the conditions in (\ref{K(0)K(1)}) become $\kappa(\infty)=0$ and $\kappa(0)=1$, respectively.
The affinity-based extended Tsallis entropy (\ref{Tsallis_d}) is rewritten in terms of the distance variable as
\begin{align}\label{Tsallis_rho}
\cH_{q}(\{p_{i}\},\{\rho_{ij}\})
&=\ln_{q}{}^{q}\!\iD(\{p_{i}\},\{\rho_{ij}\})\no\\
&=\begin{cases}
\, \ds \f{1}{1-q}\sum_{\mathrm{all}\, i}p_{i}\Bigg[\Bigg(\sum_{\mathrm{all}\, j}p_{j}\kappa(\rho_{ij})\Bigg)^{q-1}-1\Bigg] & (q\neq 1) \\[4mm]
\, \ds -\sum_{\mathrm{all}\, i}p_{i}\ln\Bigg( \sum_{\mathrm{all}\, j}p_{j}\kappa(\rho_{ij}) \Bigg)  & (q=1)
\end{cases}\,.
\end{align}
Here and hereafter we set $P=\sum_{\mathrm{all}\, i}p_{i}=1$ for simplicity (without loss of generality).

\subsection{Basic relations and properties}
Let us first investigate the pseudo-additivity of the extended entropy.
We consider the affinity-based extended entropy of the combined system $\iS_{IJ}$ composed of two systems $\iS_{I}$ and $\iS_{J}$.
Here the entropy of $\iS_{I}$ is given by a function of $\{p_{i_{I}}\}$ and $\{\rho_{i_{I}j_{I}}\}$, where $i_{I},\,j_{I}\in\cS_{I}=\{1,\,\dots,\,\,n_{I}\}$ with $\sum_{i_{I}=1}^{n_{I}}p_{i_{I}}=1$, and likewise for $\iS_{J}$.
If the two systems are independent in the sense that the probability distribution of $\iS_{IJ}$ factorises into those of $\iS_{I}$ and $\iS_{J}$, then the following multiplicative relation between the effective diversities (i.e.\ the effective numbers of maximally dissimilar states) holds by definition:
\begin{equation}\label{multiplicative}
{}^{q}\!\iD(\iS_{IJ})={}^{q}\!\iD(\iS_{I})\cdot{}^{q}\!\iD(\iS_{J})\,.
\end{equation}
This relation translates to, in the light of $\cH_{q}=\ln_{q}{}^{q}\!\iD$,
\begin{align}\label{pseudo-additivity}
\cH_{q}(\iS_{IJ})
=\cH_{q}(\iS_{I})+\cH_{q}(\iS_{J})
+(1-q)\cH_{q}(\iS_{I})\cH_{q}(\iS_{J})\,,
\end{align}
which manifests the pseudo-additivity for the current extended non-extensive entropy case, with $\cH_{q}(\iS_{IJ})$ the extended joint entropy of $\iS_{I}$ and $\iS_{J}$.
The order parameter $q$, which controlled the sensitivity to common or rare categories in diversity evaluation, now takes on the meaning of the degree of non-extensivity, just as in the standard Tsallis statistics case.
This relation readily generalises to cases with more than two ($N>2$) subsystems.
The condition of independence of the systems, ${}^{q}\!\iD(\iS_{1,\,\dots,\,N})=\prod_{I=1}^{N}{}^{q}\!\iD(\iS_{I})$, translates to the following pseudo-additivity relation for the $N$-combined system:
\begin{align}
\cH_{q}(\iS_{1,\,\dots,\,N})&=\sum_{I=1}^{N}\cH_{q}(\iS_{I})
+(1-q)\sum_{I<I'}\cH_{q}(\iS_{I})\cH_{q}(\iS_{I'})
+(1-q)^{2}\sum_{I<I'<I''}\cH_{q}(\iS_{I})\cH_{q}(\iS_{I'})\cH_{q}(\iS_{I''})+\cdots\no\\
&=\sum_{L=1}^{N}(1-q)^{L-1}\Bigg[\sum_{I_{1}<\dots<I_{L}}\prod_{\ell=1}^{L}\cH_{q}(\iS_{I_{\ell}})\Bigg]\,.
\end{align}
Introducing the shorthand notations $i(N)\coloneqq (i_{1},\,\dots,\,i_{N})$ and $\sum_{i(N)}\coloneqq\sum_{i_{1}=1}^{n_{1}}\cdots\sum_{i_{N}=1}^{n_{N}}$, it can be expressed as
\begin{align}
&\cH_{q}(\iS_{1,\,\dots,\,N})
=\f{1}{1-q}\Bigg[\sum_{i(N)}p_{i(N)}\Bigg(\sum_{j(N)}p_{j(N)}\kappa\big(\rho_{i(N);\,j(N)}\big)\Bigg)^{q-1}-1\Bigg]\,,\label{N_system}\\
&\hspace{3mm}\text{where}\quad p_{i(N)}\coloneqq\prod_{I=1}^{N}p_{i_{I}}\quad \text{and}\quad
\kappa\big(\rho_{i(N);\,j(N)}\big)\coloneqq\prod_{I=1}^{N}\kappa(\rho_{i_{I}j_{I}})\,,
\quad \text{or}\quad \rho_{i(N);\,j(N)}=\kko{\sum_{I=1}^{N}\ko{\rho_{i_{I}j_{I}}}^{\eta}}^{1/\eta}\,.\no
\end{align}
As usual, the extensive statistics is recovered in the $q\to 1$ limit, when only the term corresponding to $L=1$ survives.
It further reduces to Boltzmann--Gibbs entropy in the zero-affinity limit: $\kappa(\rho_{ij})\to \delta_{ij}$.
For all the above properties, see \ref{app:divergence} for a more general discussion that outlines some information-theoretic properties of the current extended framework.

A comment on the concavity of the affinity-based extended entropy concerning general probability distribution follows.
The Hessian components of the entropy function is computed as
\begin{align}
\nabla^{2}\cH_{q}&=\f{\pa^{2}}{\pa{p_k}\pa{p_{\ell}}}\cH_{q}(\{p_{i}\},\{\rho_{ij}\})\no\\
&{}=-\kko{\Bigg(\sum_{j=1}^{n}p_{j}\kappa(\rho_{kj})\Bigg)^{q-2}+\Bigg(\sum_{j=1}^{n}p_{j}\kappa(\rho_{\ell j})\Bigg)^{q-2}}\kappa(\rho_{k\ell})
-(q-2)\sum_{i=1}^{n}p_{i}\Bigg( \sum_{j=1}^{n}p_{j}\kappa(\rho_{ij})\Bigg)^{q-3}\kappa(\rho_{ik})\kappa(\rho_{i\ell})\,.
\end{align}
From this expression, it can be verified that the Hessian is negative-definite if $q\geq 2$, when the entropy function becomes strictly concave.
It is also easily seen that, in the Tsallis limit, $\kappa(\rho_{ij})\to \delta_{ij}$, the above Hessian components reduces to $-q(p_{k})^{q-2}\delta_{kl}\, (<0)$, indicating the known result that the Tsallis entropy function is strictly concave with respect to $\{p_{i}\}$ for all $q>0$.

\subsection{Microcanonical ensemble}

The Jaynes maximum entropy principle \cite{Jaynes83} states that the states of thermodynamic equilibrium are states of maximum entropy.
Here we investigate its implications on the affinity-based extended non-extensive entropy.
We could call the resulting guiding principle the \q{maximum effective diversity principle}.
Let us consider a closed Hamiltonian system at energy $E$, whose Lagrangian for the microcanonical ensemble is given by
\begin{align}
\cL_{q}[\{p_{i}\},\{\rho_{ij}\};\alpha]=\cH_{q}(\{p_{i}\},\{\rho_{ij}\})-\alpha\Bigg(\sum_{i=1}^{n}p_{i}-1\Bigg)\,,
\end{align}
where $\cH_{q}$ is the extended entropy of (\ref{Tsallis_rho}), $p_{i}$ denotes the probability for each of the $n(E)$ microscopic configurations labelled by $i=1,\,\dots,\,n$, and $\alpha$ is the Lagrange multiplier for the probability conservation constraint.
Following the standard procedure, necessary conditions that the set $\{p_{k}\}$ represents the maximum entropy state are $\pa\cL/\pa{p_{k}}=0$ for $k=1,\,\dots,\,n$, and $\pa\cL/\pa{\alpha}=0$.
Using these conditions, we obtain
\begin{align}\label{mod.eq.prob}
\alpha=\f{1}{1-q}\Bigg[
\Bigg(\sum_{j=1}^{n}p_{j}\kappa(\rho_{kj})\Bigg)^{q-1}-1\Bigg]
-\sum_{i=1}^{n}p_{i}\kappa(\rho_{ik})\Bigg(\sum_{j=1}^{n}p_{j}\kappa(\rho_{ij})\Bigg)^{q-2}
\end{align}
for $k=1,\,\dots,\,n$.
If the between-state distances, $\{\rho_{ij}\}$, are set to be constant across all $i\neq j$, then the above relation implies that $p_{k}$ is independent of $k$.
This, upon imposition of the normalisation condition $\pa\cL/\pa{\alpha}=0$, immediately leads to $p_{k}=1/n(E)$ for all $k$, i.e.\ the entropy attains its maximum for the uniform distribution.
In this case, it follows that $\cH_{q}^{\max}=\cH_{q}(\{1/n(E)\})=\ln_{q}n(E)$, which is the $q$-deformed version of the Boltzmann's equation (\ref{Boltzmann}).
However, with a set of nontrivial distances $\{\rho_{ij}\}$, the extended entropy is no longer maximised by the uniform distribution, and therefore, the postulate of equal {a priori} probabilities must be modified in accordance with the new constraint (\ref{mod.eq.prob}).

\begin{figure}[tp]
\centering
\includegraphics[width=14.5cm]{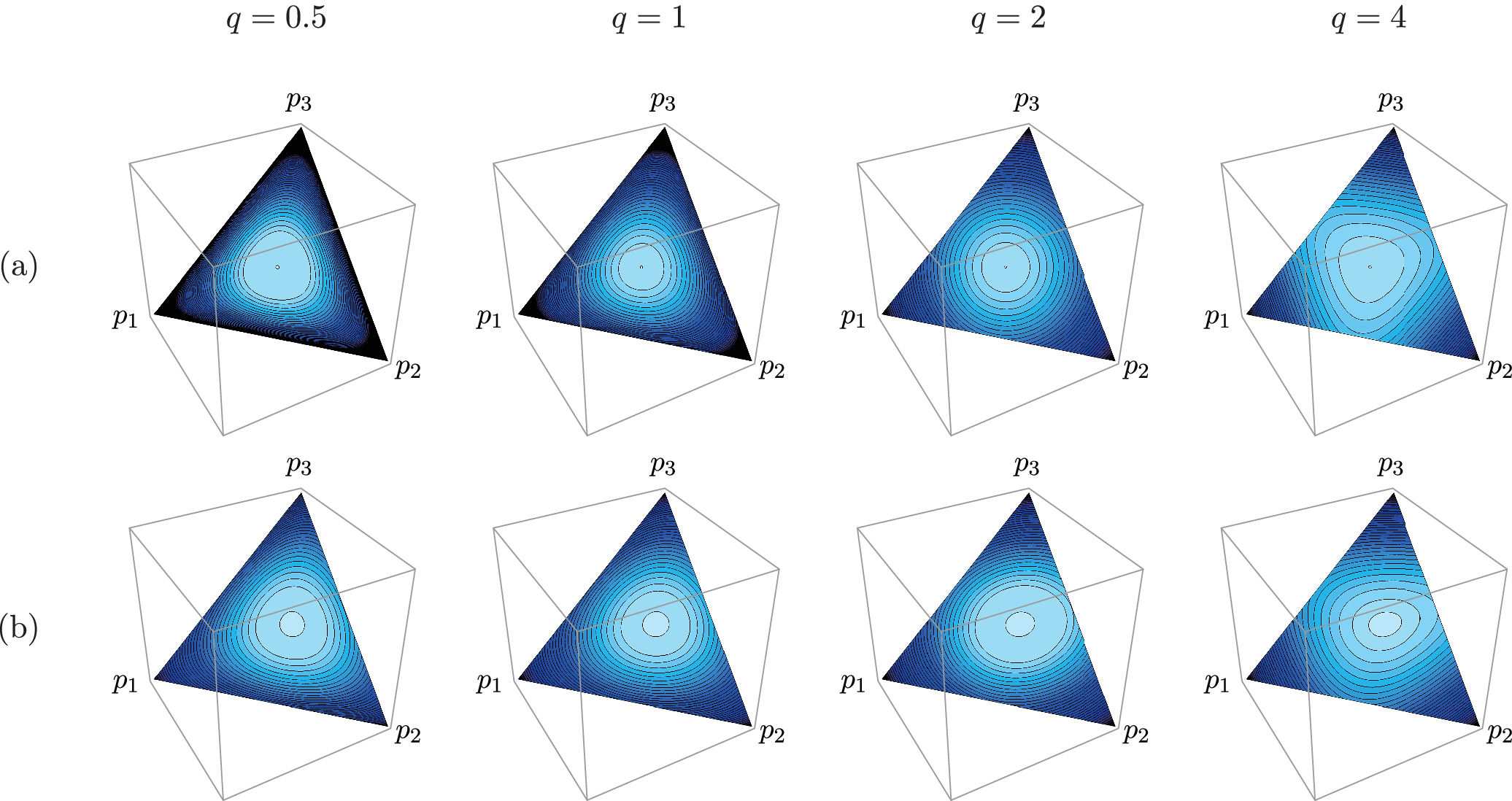}
\caption{Contour plots of the entropy function $\cH_{q}$ for fixed $\{d_{ij}\}$, projected onto the plane defined by $\sum_{i}p_{i}=1$ with $p_{i}\in[0,\,1]$ for $i=1,\,2,\,3$, where the brighter (resp.\ darker) region indicates larger (resp.\ smaller) values of the entropy.
Contour plots corresponding to $q=0.5,\,1,\,2,\,4$ cases are shown for each of $\mathrm{(a)}$ Tsallis entropy ($d_{ij}=1$ for all distinct pairs of $i$ and $j$) and $\mathrm{(b)}$ the affinity-based extended Tsallis entropy ($d_{12}^{*}=0.340$, $d_{13}^{*}=0.263$ and $d_{23}^{*}=0.0776$).}
\label{fig:maxent}
\end{figure}

A graphical illustration is useful here.
Fig.\ \ref{fig:maxent} shows contour plots of the extended Tsallis entropy $\cH_{q}$ as functions of $\{p_{i}\}$ for a fixed set of $\{d_{ij}\}$,\footnote{Here we employ the original dissimilarity variables $\{d_{ij}\}$ instead of the distance variables $\{\rho_{ij}\}$, and also set $r=1$ for simplicity.} projected onto the plane defined by $\sum_{i}p_{i}=1$ with $i,\,j=1,\,2,\,3$ (i.e.\ $n=3$ case).
The simulation results are shown for different values of the order parameter (i.e.\ the degree of non-extensivity): $q=0.5,\,1,\,2,\,4$.
The four diagrams in Fig.\ \ref{fig:maxent}(a) correspond to the Tsallis entropy case ($d_{ij}=1-\delta_{ij}$), when the postulate of equal {a priori} probabilities is satisfied; i.e.\ the entropy is maximised at $p_{i}^{\mathrm{(eq)}}=1/3$ for $i=1,\,2,\,3$ regardless of the value of $q$.
The four diagrams in Fig.\ \ref{fig:maxent}(b) correspond to the affinity-based extended Tsallis entropy case with the dissimilarity variables given by $d_{12}^{*}=0.340$, $d_{13}^{*}=0.263$ and $d_{23}^{*}=0.0776$.
As can be seen, in the presence of nontrivial between-state affinities, the entropy is no longer maximised at the previous equiprobable point but at a certain shifted point defined by the conditions (\ref{mod.eq.prob}).
For example, in the $q=2$ case, the entropy is maximised at $(p_{1}^{*}{}^{\mathrm{(eq)}},\,p_{2}^{*}{}^{\mathrm{(eq)}},\,p_{3}^{*}{}^{\mathrm{(eq)}})\approx (0.254,\,0.362,\,0.383)$, giving $\cH_{2}^{\max}(\{p_{i}^{*}{}^{\mathrm{(eq)}}\},\{d_{ij}^{*}\})\approx 0.521$, which is lower than that computed for the Tsallis statistics case: $\cH_{2}^{\max}(\{p_{i}^{\mathrm{(eq)}}=1/3\},\{d_{ij}=1-\delta_{ij}\})=2/3$.

\subsection{Canonical ensemble}

Next, let us demonstrate how the canonical distribution is derived from the affinity-based extended Tsallis entropy.
Here we focus on the $q=2$ case, when the system respects the Nesting Principle.
This means that the system's effective diversity, i.e.\ the effective number of maximally dissimilar microstates, does not change even if the system is coarse-grained or become \q{degenerated} in the affinity dimension depending on the scale of observation.
Here, due to the quadratic nature of the system, the following vector--matrix notation is useful:
\begin{equation}
\bs{A}\coloneqq[\kappa(\rho_{ij})]\,,\quad
\ket{\bs{p}}\coloneqq(p_{1},\,\dots,\,p_{n})^{\intercal}\,,
\end{equation}
where $\bs{A}$ is an $n\times n$ symmetric matrix (\q{affinity matrix}) whose $(i,j)$-component is given by $\kappa(\rho_{ij})$, and $\bs{p}$ is an $n$-vector whose $i$-th component is given by $p_{i}$.
With $\{\ket{\bold{e}_{i}}\}_{i=1}^{n}$ denoting the unit vector basis of $\mathbb{R}^{n}$, the distribution vector can also be written as $\ket{\bs{p}}=\sum_{i=1}^{n}p_{i}\ket{\bold{e}_{i}}$.
For later notational convenience, we also define $\ket{\bs{c}}\coloneqq\sum_{i=1}^{n}\ket{\bold{e}_{i}}$.
The inner product of two vectors $\ket{\bs{v}}$ and $\ket{\bs{u}}$ is given by $\braket{\bs{u}|\bs{v}}=\sum_{i=1}^{n}u_{i}v_{i}$, and the matrix ($\bs{M}$) operation on a vector ($\bs{v}$) is computed as $\bs{M}\ket{\bs{v}}=\sum_{i=1}^{n}\sum_{j=1}^{n}M_{ij}v_{j}\ket{\bold{e}_{i}}$.

Our goal is to obtain the equilibrium canonical distribution, $\ket{\bs{p}^{\mathrm{(eq)}}}$, subject to a constraint on a physical quantity $\bs{Q}$ (e.g.\ Hamiltonian of the system) that takes either value of $Q_{i}$ ($i=1,\,\dots,\,n$), this time in the presence of general between-state affinities.
Caution has to be taken in defining the expectation of $\bs{Q}$.
Recall that, in the Tsallis $q$-statistics case \cite{Tsallis98} already, it is defined neither by $\sum_{i=1}^{n}p_{i}Q_{i}$ nor $\sum_{i=1}^{n}(p_{i})^{q}Q_{i}$ but by $\sum_{i=1}^{n}\pi_{i}Q_{i}$ with $\pi_{i}\coloneqq (p_{i})^{q}/\sum_{j=1}^{n}(p_{j})^{q}$ the so-called escort distribution, satisfying $\sum_{i=1}^{n}\pi_{i}=1$.
Here, we adopt the following definition for the expectation of $\bs{Q}$ in our affinity-based extended $q=2$ Tsallis statistics:\footnote{It is assumed that the degeneracy of the spectrum of $\bs{Q}$ is properly taken into account in the definition of (\ref{def:vev}), so that a modified version of the postulate of equal {a priori} probabilities compatible with the constraint (\ref{mod.eq.prob}) is satisfied.}
\begin{equation}\label{def:vev}
\vev{\bs{Q}}\coloneqq \f{\bra{\bs{p}}\bs{QA}\ket{\bs{p}}}{\bra{\bs{p}}\bs{A}\ket{\bs{p}}}\,,\quad
\bs{Q}\coloneqq \diag(Q_{1},\,\dots,\,Q_{n})
\end{equation}
so that it reproduces the $q=2$ Tsallis statistics case in the zero-affinity limit $\kappa(\rho_{ij})\to\delta_{ij}$, i.e.\ in the limit $\bs{A}$ approaches the $n$-dimensional identity matrix $\bs{I}_{n}$.
Subsequently, the Lagrangian of the system can be written as
\begin{align}
\cL[\bs{p},\bs{A};\alpha,\beta]= \cH(\bs{p},\bs{A})-\alpha\ko{\braket{\bs{c}|\bs{p}}-1}
-\beta\ko{\f{\bra{\bs{p}}\bs{QA}\ket{\bs{p}}}{\bra{\bs{p}}\bs{A}\ket{\bs{p}}}-\vev{\bs{Q}}}\,,
\end{align}
where the Hamiltonian is given by
\begin{equation}
\cH(\bs{p},\bs{A})=\braket{\bs{c}|\bs{p}}-\bra{\bs{p}}\bs{A}\ket{\bs{p}}\,,
\end{equation}
and the Lagrange multipliers $\alpha$ and $\beta$ are associated with the probability conservation condition $\sum_{i=1}^{n}p_{i}=1$ and the expectation condition (\ref{def:vev}), respectively.
Following the standard procedure, the conditions $\pa{\cL}/\pa{\alpha}=0$ and $\pa{\cL}/\pa{\beta}=0$ respectively lead to
\begin{equation}\label{Lag23}
\braket{\bs{c}|\bs{p}}=1\quad\text{and}\quad\bra{\bs{p}}\widetilde{\bs{Q}}\bs{A}\ket{\bs{p}}=0\,,\quad
\widetilde{\bs{Q}}\coloneqq \bs{Q}-\vev{\bs{Q}}\bs{I}_{n}\,,
\end{equation}
and the condition $\pa{\cL}/\pa{\ket{\bs{p}}}=0$ leads to
\begin{align}\label{Lag1}
-2\bs{A}\ket{\bs{p}}+(1-\alpha)\ket{\bs{c}}-\beta\f{\{\bs{A},\widetilde{\bs{Q}}\}\ket{\bs{p}}}{\bra{\bs{p}}\bs{A}\ket{\bs{p}}}=\ket{\bs{0}}\,,
\quad \ds \{\bs{A},\widetilde{\bs{Q}}\}\coloneqq \bs{A}\widetilde{\bs{Q}}+\widetilde{\bs{Q}}\bs{A}\,.
\end{align}
By evaluating the scalar product of $\bra{\bs{p}}$ and (\ref{Lag1}) and using (\ref{Lag23}), we obtain the relation
\begin{equation}
\bra{\bs{p}}\bs{A}\ket{\bs{p}}=\mathcal{Z}^{-1}\,,
\quad \cZ\coloneqq \f{2}{1-\alpha}\,,
\end{equation}
where the \q{partition function} $\cZ$ equals the effective diversity of the system.
The condition (\ref{Lag1}) can then be cast into the form
\begin{equation}\label{def_R}
\bs{R}\ket{\bs{p}}=\cZ^{-1}\ket{\bs{c}}\,,\quad
\bs{R}\coloneqq \bs{A}+\f{\beta\cZ}{2}\{\bs{A},\widetilde{\bs{Q}}\}\,.
\end{equation}
Assuming the invertibility of the above-defined symmetric matrix $\bs{R}$, this equation is solved to give the canonical distribution,
\begin{equation}\label{canonical}
\ket{\bs{p}^{\mathrm{(eq)}}}=\cZ^{-1}\bs{R}^{-1}\ket{\bs{c}}\,,
\end{equation}
along with the maximum entropy at equilibrium,
\begin{align}
\cH^{\mathrm{(eq)}}
\coloneqq \cH (\bs{p}^{\mathrm{(eq)}},\bs{A})
=\ln_{q=2}\cZ^{\mathrm{(eq)}}(\bs{p}^{\mathrm{(eq)}},\bs{A})
=1-\bra{\bs{p}^{\mathrm{(eq)}}}\bs{A}\ket{\bs{p}^{\mathrm{(eq)}}}\,.
\end{align}
Note that, here, the physicality conditions for $\ket{{\bs{p}}^{\mathrm{(eq)}}}$ (i.e.\ $p_{i}^{\mathrm{(eq)}}\in [0,1]$ for all $i$, subject to $\sum_{i=1}^{n}p_{i}^{\mathrm{(eq)}}=1$) impose constraints on the possible relationships of the Lagrange multipliers, given $\bs{A}$ and $\bs{Q}$.
Moreover, by taking $\bs{Q}$ to be the Hamiltonian of the system, the expression for the physical temperature \cite{Abe01,Suyari06} can be obtained.
Writing $\vev{\bs{Q}}\mapsto U$, a small computation leads to:
\begin{align}
\f{\pa{\cH^{\mathrm{(eq)}}}}{\pa{U}}
=\ko{\cZ^{\mathrm{(eq)}}}^{-2}\f{\pa \cZ^{\mathrm{(eq)}}}{\pa U}
=\ko{\cZ^{\mathrm{(eq)}}}^{-2}\Braket{\bs{c}|\f{\pa{\bs{R}^{-1}}}{\pa{U}}|\bs{c}}
=-\Braket{\bs{p}^{\mathrm{(eq)}}|\f{\pa{\bs{R}}}{\pa{U}}|\bs{p}^{\mathrm{(eq)}}}\,,
\end{align}
where we have used Eq.\ (\ref{canonical}) and the relation $\pa \bs{R}^{-1}/\pa U=-\bs{R}^{-1}(\pa \bs{R}/\pa U)\bs{R}^{-1}$.
By using the definition of $\bs{R}$ in (\ref{def_R}) and also noticing that $\{\bs{A},\widetilde{\bs{Q}}\}$ vanishes by sandwiching it between $\bra{\bs{p}^{\mathrm{(eq)}}}$ and $\ket{\bs{p}^{\mathrm{(eq)}}}$, which is due to the second constraint in (\ref{Lag23}), the above expression further reduces to
\begin{align}\label{def:T}
\f{\pa{\cH^{\mathrm{(eq)}}}}{\pa{U}}
=\beta\cZ^{\mathrm{(eq)}}\braket{\bs{p}^{\mathrm{(eq)}}|\bs{A}|\bs{p}^{\mathrm{(eq)}}}
=\beta_{2}\big(1-\cH^{\mathrm{(eq)}}\big)\,,\quad \beta_{2}\coloneqq \f{2\beta}{1-\alpha}\,.
\end{align}
Subsequently, the physical temperature of the $q=2$ affinity-based extended Tsallis system is identified to be, based on the same reasoning as the standard Tsallis case \cite{Abe01,Suyari06},
\begin{equation}
T\coloneqq \beta_{2}^{-1}=\big(1-\cH^{\mathrm{(eq)}}\big)\ko{\f{\pa{\cH^{\mathrm{(eq)}}}}{\pa{U}}}^{-1}\,,
\end{equation}
with which the generalised thermodynamic quantities compatible with the thermodynamic Legendre transform structure are obtained.
For instance, the free energy and the specific heat are calculated to be $F=U+T\ln\ko{1-\cH^{\mathrm{(eq)}}}$ and $C_{\mathrm{V}}=\pa U/\pa T=-T\pa^{2} F/\pa T^{2}$, respectively.
All the above results generalise the known results \cite{Abe01,Wada05,Suyari06} for the $q=2$ Tsallis statistical mechanics, which are reproduced as the zero-affinity limit $\bs{A}\to \bs{I}_{n}$ of the above results.
It is also straightforward to generalise the above results to cases with multiple constraints $\braket{\bs{p}|\bs{Q}^{(\ell)}\bs{A}|\bs{p}}\braket{\bs{p}|\bs{A}|\bs{p}}^{-1}=\langle\bs{Q}^{(\ell)}\rangle$, $\ell=1,\,\dots,\,m$; one only needs to introduce as many Lagrange multiplies $\{\beta_{1},\,\dots,\,\beta_{m}\}$ in the formulation.
The results can also be straightforwardly extensible to the general $q$ case.

\subsection{Application: A two-level system with affinity}

\begin{figure}[tp]
\centering
\includegraphics[width=0.85\linewidth]{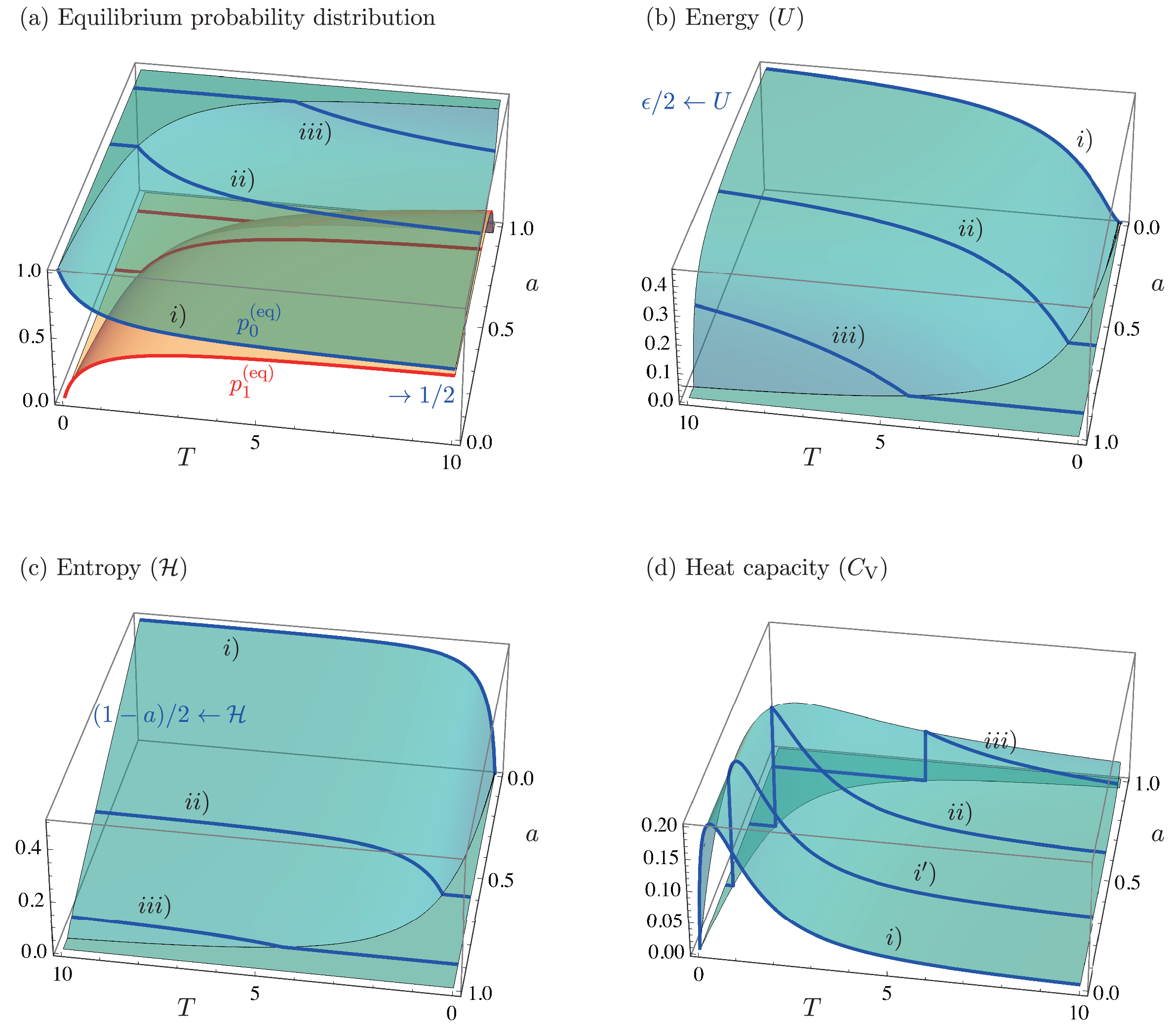}
\vspace{6mm}
\caption{Temperature ($T$) dependence of the thermodynamic quantities at equilibrium depicted for a two-level system with energy gap $\ep=1$ and between-microstate affinity $a\in [0,\,1)$.
Each diagram represents: (a) probability distribution, $(p_{0}^{\mathrm{(eq)}},\,p_{1}^{\mathrm{(eq)}})$, where the upper blue line and the lower red line correspond to, respectively, $p_{0}^{\mathrm{(eq)}}$ and $p_{1}^{\mathrm{(eq)}}$ (the surface for $p_{0}^{\mathrm{(eq)}}$ and that for $p_{1}^{\mathrm{(eq)}}$ are symmetric with respect to the horizontal plane defined by $p=1/2$), (b) energy, $U$, (c) entropy, $\cH$, and (d) heat capacity, $C_{\mathrm{V}}$.  
The depicted curves in each diagram correspond to different values of the affinity: $i)$ $a=0$, $ii)$ $a=0.6$ and $iii)$ $a=0.9$, respectively; the curve corresponding to $i')$ $a=0.3$ is additionally depicted for (d).
Below the critical temperature $T_{0}=(a\ep/2)(1-a)^{-1}$, the thermodynamic quantities ($U$, $\cH$ and $C_{\mathrm{V}}$) remain zero as the probability distribution is fixed to $(p_{0}^{\mathrm{(eq)}},\,p_{1}^{\mathrm{(eq)}})=(1,\,0)$.
Note that the positive directions of the $T$- and $a$-axes of (b) and (c) are reversed from those of (a) and (d) for ease of description.}
\label{fig:thermbehaviour}
\end{figure}

Based on the above results, let us finally explore how the thermodynamic quantities are affected in the presence of between-state affinities by means of a concrete model.
As an illustration, we apply the above $q=2$ extended canonical model to a two-level system (non-degenerate), where $p_{0}$ and $p_{1}$ represent the probability of a particle being found in the ground level with energy $0$ and in the excited level with energy $\ep\,(>0)$, respectively.
The new characteristic of the system compared to the Tsallis or Boltzmann--Gibbs case is the between-microstate affinity, denoted by $a\coloneqq\kappa(\rho_{01})\in [0,\,1)$.
Using the same notation as before, the energy of the system is defined and calculated to be
\begin{equation}
U= \f{\bra{\bs{p}}\bs{UA}\ket{\bs{p}}}{\bra{\bs{p}}\bs{A}\ket{\bs{p}}}\,,\quad
\text{where}\quad
\ket{\bs{p}}=\begin{pmatrix}
p_{0} \\
p_{1} \\
\end{pmatrix}\,,\quad 
\bs{A}=\begin{pmatrix}
1 & a \\
a & 1 \\
\end{pmatrix}\,,\quad
\bs{U}=\begin{pmatrix}
0 & 0 \\
0 & \ep \\
\end{pmatrix}\,. 
\end{equation}
The equilibrium distribution $\ket{\bs{p}^{\mathrm{(eq)}}}$ is obtained by solving Eq.\ (\ref{canonical}) with the Lagrange multiplier constraints (\ref{Lag23}).
The set of equations can be solved analytically; the resulting expressions for the probability distribution, the energy, the entropy and the heat capacity of the system at equilibrium are given by, respectively,
\begin{align}
&\ket{\bs{p}^{\mathrm{(eq)}}}=\begin{pmatrix}
p_{0}^{\mathrm{(eq)}} \\
p_{1}^{\mathrm{(eq)}} \\
\end{pmatrix}\,,\quad
p_{0}^{\mathrm{(eq)}}=\f{1}{2}+\f{3(1+a)Y^{1/3}\ep}{2W}\,,\quad 
p_{1}^{\mathrm{(eq)}}=1-p_{0}^{\mathrm{(eq)}}\,,\\[2mm]
&U
=\f{\braket{\bs{p}^{\mathrm{(eq)}}|\bs{UA}|\bs{p}^{\mathrm{(eq)}}}}{\braket{\bs{p}^{\mathrm{(eq)}}|\bs{A}|\bs{p}^{\mathrm{(eq)}}}}
=\f{\ep}{2}-\f{3Y^{1/3}W\ep^{2}}{9XY^{2/3}\ep^{2}+W^{2}}\,,\\[2mm]
&\cH
=1-\braket{\bs{p}^{\mathrm{(eq)}}|\bs{A}|\bs{p}^{\mathrm{(eq)}}}
=\f{1-a}{2}\ko{1-\f{9(1+a)^{2}Y^{2/3}\ep^{2}}{W^{2}}}\,,\\[2mm] 
&C_{\mathrm{V}}
=\f{\pa U}{\pa T}
=-2T\f{\braket{\bs{p}^{\mathrm{(eq)}}|\bs{A}|\pa_{T}\bs{p}^{\mathrm{(eq)}}}}{\braket{\bs{p}^{\mathrm{(eq)}}|\bs{A}|\bs{p}^{\mathrm{(eq)}}}}=\f{2}{3}-\f{4XT}{3Y^{1/3}}-\f{\pa_{T}W}{18XY^{2/3}}+\f{\big(4XT^{2}+3\ep^{2}\big)\pa_{T}W}{18Y^{4/3}}\,,\quad 
\pa_{T}\coloneqq \pa/\pa T\,,\label{heat capacity}\\[2mm]
&\text{where}\quad 
X\coloneqq1-a^{2}\,,\quad 
Y\coloneqq8X^{3}T^{3}+36X^{2}T\ep^{2}+3\ep\sqrt{3X^{3}\big( 16X^{2}T^{4}+44XT^{2}\ep^{2}-\ep^{4} \big)}\,,\no\\[0mm]
&\quad \text{and}\quad W\coloneqq 4X^{2}T^{2}+2XY^{1/3}T+Y^{2/3}+3X\ep^{2}\,.\no
\end{align}
Here, the physicality conditions for $\ket{\bs{p}^{\mathrm{(eq)}}}$ (i.e.\ $p_{0}^{\mathrm{(eq)}},\,p_{1}^{\mathrm{(eq)}}\in [0,\,1]$ and $p_{0}^{\mathrm{(eq)}}+p_{1}^{\mathrm{(eq)}}=1$) require that there exists a critical temperature below which the particle is always found in the ground level (i.e.\ $p_{0}^{\mathrm{(eq)}}=1$);
this characteristic temperature is given by
\begin{equation}\label{freezing}
T_{0}= \f{a\ep}{2(1-a)}\,.
\end{equation}
Note that if the system consisted of a macroscopic number $N$ of distinguishable particles (e.g.\ atoms in a solid), in which $Np_{0}$ and $Np_{1}$ represent the population of the ground level and the excited level, respectively, then the consequence of the above physicality condition is that all $N$ particles \q{freeze} to the ground level for the low-temperature region $[0,\,T_{0}]$.
It is noteworthy that this phenomenon is neither due to attractive interactions between the particles nor the Bose--Einstein type condensation but purely due to the between-microstate affinity.
Equation (\ref{freezing}) indicates that the \q{freezing temperature} depends on the magnitude of the affinity.
Evidently, $T_{0}\to 0$ as $a\to 0$, hence no such \q{freezing} above $T=0$ occurs in the Tsallis/Boltzmann--Gibbs case.
As the affinity increases, $T_{0}$ also increases so that the \q{freezing} persists up to high temperature.
Fig.\ \ref{fig:thermbehaviour}(a) shows the thermal behaviour of the probability distribution at equilibrium ($\bs{p}^{\mathrm{(eq)}}$).
As the temperature is increased, the probabilities $p_{0}^{\mathrm{(eq)}}$ and $p_{1}^{\mathrm{(eq)}}$ tend to the same asymptotic value of $1/2$ from above and below, respectively, exhibiting the uniform distribution in the high-temperature limit $T\to \infty$.
 
Fig.\ \ref{fig:thermbehaviour}(b) shows the thermal behaviour of the equilibrium energy ($U$).
There is a zero-energy plateau for the low-temperature region $[0,\,T_{0}]$, reflecting the fact that in that region, the probability of a particle being found in the excited level is zero.
Above $T_{0}$ the energy increases monotonically with the temperature and tends to the half-energy gap $\ep/2$ as $T\to \infty$.
In addition, Fig.\ \ref{fig:thermbehaviour}(c) shows the thermal behaviour of the equilibrium entropy ($\cH$).
There is a zero-entropy plateau for $[0,\,T_{0}]$, reflecting again the fact that the effective diversity of the system is one (i.e.\ single-microstate concentration), also satisfying the third law of thermodynamics.
The fact that the entropy vanishes for $[0,\,T_{0}]$ also indicates that the physical temperature variable agrees between the extensive case ($q=1$) and the current non-extensive case ($q=2$) for the low-temperature region, in the light of $T_{q}=\big(1+(1-q)\cH\big)T_{q=1}$ \cite{Abe01}.
Above the critical temperature $T_{0}$, the effective diversity increases monotonically with the temperature, which tends to $\iD\to 2/(1+a)$ as $T\to \infty$, and hence the entropy to $\cH=\ln_{q=2}\iD\to (1-a)/2$.

Finally, Fig.\ \ref{fig:thermbehaviour}(d) shows the thermal behaviour of the equilibrium heat capacity at constant volume ($C_{\mathrm{V}}$).
In the zero-affinity case ($a=0$; curve $i$), its behaviour resembles the Schottky anomaly of the extensive ($q=1$) case, which is typical for the two-level systems; it increases from zero at $T=0$ through a maximum at around $T\sim \mathcal{O}(\ep)$ and subsequently falls to zero as $T\to\infty$.
However, with non-vanishing affinity, the heat capacity exhibits a gap of magnitude $C_{\mathrm{V}}(T_{0})$ at the critical temperature $T_{0}$.
Here, another characteristic temperature $T_{1}$ of the system is of importance, at which the heat capacity function $C_{\mathrm{V}}(T)$, given by (\ref{heat capacity}), reaches its maximum peak ($C_{\mathrm{V}}(T_{1})=0.2$), when the effective diversity of the system equals $\iD=(3/2)/(1+a)$.
Whether such a peak is observed physically depends on the magnitude of the affinity, and accordingly, there are two patterns of the thermal behaviour of the heat capacity above $T_{0}$.
If $a<1/2$ (i.e.\ low affinity; curve $i'$), for which $T_{1}>T_{0}$, then the heat capacity increases to its maximum at $T_{1}$ before monotonically decreasing to zero as $T\to\infty$; while if $a\geq 1/2$ (i.e.\ high affinity; curves $ii$ and $iii$), for which $T_{1}\leq T_{0}$, then the heat capacity monotonically decreases from $C_{\mathrm{V}}(T_{0})$ to zero as $T\to \infty$.

It would be interesting if the new phenomena described above, including the emergence of the \q{freezing} phase below the critical temperature $T_{0}$, could actually be observed experimentally.
For instance, with sufficient precision of the measurement of $T_{0}$ (possibly through the investigation of the thermal behaviour of the heat capacity) and the energy gap $\ep$, the between-state affinity could be detected and estimated by the relation $a=2T_{0}/(2T_{0}+\ep)$, which behaves as $a\sim 2T_{0}/\ep$ for $T_{0}/\ep\ll 1$ (low affinity) and $a\sim 1-(2T_{0}/\ep)^{-1}$ for $T_{0}/\ep\gg 1$ (high affinity).

\section{Summary and Discussions\label{sec:summary}}

In this paper, Tsallis entropy was extended to incorporate the dependence on between-microstate affinities, and subsequently, its implications for the non-extensive statistical mechanics were investigated.
At the core of our construction of the extended entropy ($\cH_{q}$) was the concept of the effective number of dissimilar microstates, termed the effective diversity (${}^{q}\!\iD$).
It was derived from the probability distribution among states, $\{p_{i}\}$, and the dissimilarities between states, $\{d_{ij}\}$, with the order parameter $q$ controlling the measure's sensitivity to the probability distribution.
The effective diversity was quantified as the number of equally-probable and maximally distinct microstates that would yield the same diversity as that of the observed system whose microstates are generally non-equally probable and could share some similarities.
The affinity-based extended Tsallis entropy was obtained from the effective diversity through the Boltzmann's-equation-like relation, $\cH_{q}=\ln_{q}{}^{q}\!\iD$, in terms of the Tsallis' $q$-logarithm, as shown in Eq.\ (\ref{Tsallis_d}).
In this construction, the effect of nontrivial affinities was incorporated through the terms involving $K(d_{ij})$ with $K(\cdot)$ the kernel function.

To ensure the universality of the measure, a new principle called the Nesting Principle was established.
It is a grouping invariance property, stating that the effective diversity/entropy measure remains invariant under an arbitrary (\q{fine-grained} or \q{coarse-grained}) grouping of the constituent states. 
For a system with general $\{d_{ij}\}$, an important consequence of this principle was that the order parameter was necessarily fixed to a specific value, i.e.\ $q=2$.
It was shown that the diversity measure with other values of $q$ either overestimates or underestimates the \q{true} effective diversity computed at the most elementary (\q{the finest-grained}) level.
The resulting effective diversity/entropy measure was indexed by the scaling parameter $r$, which controlled the weight on the between-state affinities.
Accordingly, the kernel function compatible with the Nesting Principle was explicitly obtained as $K_{r}(d_{ij})=1-(d_{ij})^{r}$.

The nesting-invariant diversity index for a system $I$, computed from Eq.\ (\ref{coarse-grainedD2}), offered a unique interpretation in terms of a generalised correlation sum defined within the same system $I$.
The formula for the effective dissimilarity between two systems $I$ and $J$, denoted as $\bar{d}_{IJ}$, was also provided as in Eq.\ (\ref{def:d_IJ}), which took the form of a cross-correlation sum, and was shown to be compatible with the Nesting Principle.
These formulae were shown to be applicable at any level of nesting structure: from the finest-grained states $\{i\}_{i=1}^{n}=\{1,\,\dots,\,n\}$ defined at the most fundamental level (i.e.\ $\iD_{i}=1$ and $d_{ii}=0$ for all $i$), to aggregate meta-level $\{I\}_{I=1}^{N}=\big\{\{i_{I}\}_{i_{I}=1}^{n_{I}}\big\}{}_{I=1}^{N}$, and further to more aggregate level $\{I'\}_{I'=1}^{N'}=\big\{\{I_{I'}\}_{I_{I'}=1}^{N_{I'}}\big\}{}_{I'=1}^{N'}=\big\{\big\{\{i_{I_{I'}}\}_{\dots}^{\dots}\big\}{}_{\dots}^{\dots}\big\}{}_{\dots}^{\dots}$ and so forth.
If applied to a closed finite system, the set of formulae produced the same value of the extended entropy regardless of the \q{granularity} of states, i.e.\ $\cH_{2}(\{p_{i}\},\{d_{ij}\})=\cH_{2}(\{P_{I}\},\{\bar{d}_{IJ}\})=\cH_{2}(\{P_{I'}\},\{\bar{\bar{d}}_{{I'}{J'}}\})=\dots=\cH_{2}(1,K^{-1}(\iD^{-1}))$.

Although it was shown that the Nesting Principle requires $q=2$ in general, it was also found that the Nesting Principle holds for arbitrary $q$ in a special situation in which $K(d_{ij})=\delta_{ij}$, i.e.\ the standard Tsallis and Boltzmann--Gibbs statistical mechanics case.
There are two routes to achieve this: one is by \q{exogenous} reduction, $d_{ij}\to 1-\delta_{ij}$, in which all the states are maximally distinct, and the other is by \q{endogenous} reduction, e.g.\ by taking the $r\to 0$ limit of $K_{r}(d_{ij})=1-(d_{ij})^{r}$.
In either case, the resulting diversity index is the very Hill numbers family, given in Eq.\ (\ref{Hill_num}), which is related to Tsallis entropy through the aforementioned Boltzmann's-equation-like relation.

Moreover, the implications of the affinity-based extended Tsallis entropy for the non-extensive statistical mechanical systems at equilibrium were investigated.
In particular, the microcanonical and the canonical ensembles were constructed in the presence of general between-state affinities.
It was shown that the equilibrium probability distributions are no longer given in the standard Tsallis' $q$-deformed forms but in a modified form with the new degrees of freedom offered by between-state affinities.
The classic postulate of equal {a priori} probabilities was also modified by the affinity-dependent terms.
Subsequently, some known results for Tsallis statistical mechanics, including the $q$-generalised thermodynamic quantities and their interrelationships, were generalised per the affinity-based extended framework.
As an illustration, the canonical method was used to make predictions concerning the macro-properties of a two-level system with affinity, which manifested some new phenomena stemming purely from the between-state affinity, such as the particle's \q{freezing} to the ground level in the low-temperature region.

We envision that the framework of the affinity-based extended entropy developed in this paper will offer useful tools for probing yet-to-be-uncovered aspects of statistical mechanics.
For instance, future applications might include a description of the non-equilibrium statistical mechanics in terms of some time-dependent distance variables $\{\rho_{ij}(t)\}$, with $t$ being the time variable.
As a simple toy model, suppose that the evolution of a system is governed by $\rho_{ij}(t)=\lambda (1-\delta_{ij})t$, $\lambda>0$, with which the time-dependent effective affinity is given by $\kappa_{ij}(t)=e^{-\rho_{ij}(t)}$.
Then the dynamics that the system undergoes can be regarded as \q{effective diversity relaxation} with $\lambda$ representing the inverse relaxation time.
Under this dynamics, the initial zero-entropy state with $\kappa_{ij}(t=0)=1$ evolves to eventually reach the equilibrium state at which $\kappa_{ij}(t\to\infty)=\delta_{ij}$, recovering Tsallis and Boltzmann--Gibbs statistical mechanics in the infinite future.

The proposed framework for diversity/entropy quantification will also be of use in a wide range of scientific disciplines beyond statistical mechanics.
The application will be either through the form of the extended Tsallis entropy, $\cH_{q}^{\mathsf{T}}=\ln_{q}{}^{q}\!\iD$, the extended R\'enyi entropy, $\cH_{q}^{\mathsf{R}}=\ln{}^{q}\!\iD$, or other forms including the effective diversity measure (${}^{q}\!\iD$) itself.\footnote{The affinity-based extended Tsallis and R\'enyi entropies can be written in the following unified form:
\begin{align}
\widetilde\cH_{q}(\{p_{i}\},\{d_{ij}\};\gamma)
=\f{1}{(1-q)\gamma}\kko{\ko{\sum_{i\in \cS} p_{i}\Bigg(\sum_{j\in \cS} p_{j}K(d_{ij})\Bigg)^{q-1}}^{\gamma}-1}\,,
\quad\text{so that}\quad
\widetilde\cH_{q}\big|_{\gamma=1}=\cH_{q}^{\mathsf{T}}
~~\text{and}~~
\widetilde\cH_{q}\big|_{\gamma\to 0}=\cH_{q}^{\mathsf{R}}\,.\no
\end{align}
}
The potential application areas would include, but are not limited to, information theory and statistical inference (see \ref{app:divergence}), biological/ecological studies (see \ref{app:biodiversity}), nonlinear dynamical systems,\footnote{The
affinity-based extended R\'enyi entropy can be used to
define a new estimator of chaotic invariants characterising an attractor in nonlinear dynamical systems.  The extended estimator is obtained by replacing the Heaviside function
(\q{hard-kernel}) in the definition of the correlation integral with
a \q{soft} kernel function of the form $\kappa(\rho_{ij})=e^{-\rho_{ij}}$, where $\rho_{ij}$ is proportional to the distance between two points labelled by $i$ and $j$ on the attractor.} measure theory and metric
geometry (see, e.g., \cite{Leinster16,Leinster16b}), quantum
information and field theories,\footnote{The affinity variables are
defined among the eigenstates of a quantum density matrix, with which the affinity-based extended quantum Tsallis and R\'enyi entropies of order $q$ can be defined.  Both entropies approach the (extended) von Neumann entropy
in the $q\to 1$ limit.  For instance, let $\ket{\Psi}=\sum_{i=1}^{n}c_{i}\ket{\psi_{i}}$ be a quantum state with $c_{i}\in\mathbb{C}$ and  $\{\ket{\psi_{i}}\}$ the set of eigenstates, then the effective affinity between $\ket{\psi_{i}}$ and $\ket{\psi_{j}}$ for the nesting-invariant ($q=2$) case can be defined by $K(d_{ij})=\mathrm{Re}(\chi_{ij}^2)/|\chi_{ij}|^{2}$, where $\chi_{ij}=c_{i}c_{j}^{*}$ is the coherence between the two engenstates.} as well as various areas of social
sciences (e.g.~\cite{Okamura19})---namely, wherever the notion of diversity
arises, and wherever the associated entropy plays a role.

\section*{Acknowledgements}
The author would like to thank Ryo Suzuki, Shuhei Mano and the two anonymous reviewers for their valuable comments and recommendations on the earlier version of the manuscript.
All content is the responsibility of the author and does not necessarily reflect the views of the organisation to which he belongs.

\appendix

\setcounter{figure}{0}
\setcounter{table}{0}
\section{Summary of Diversity Indices\label{app:diversity indices}}

\begin{figure}[th]
\centering
\includegraphics[width=13.5cm,clip]{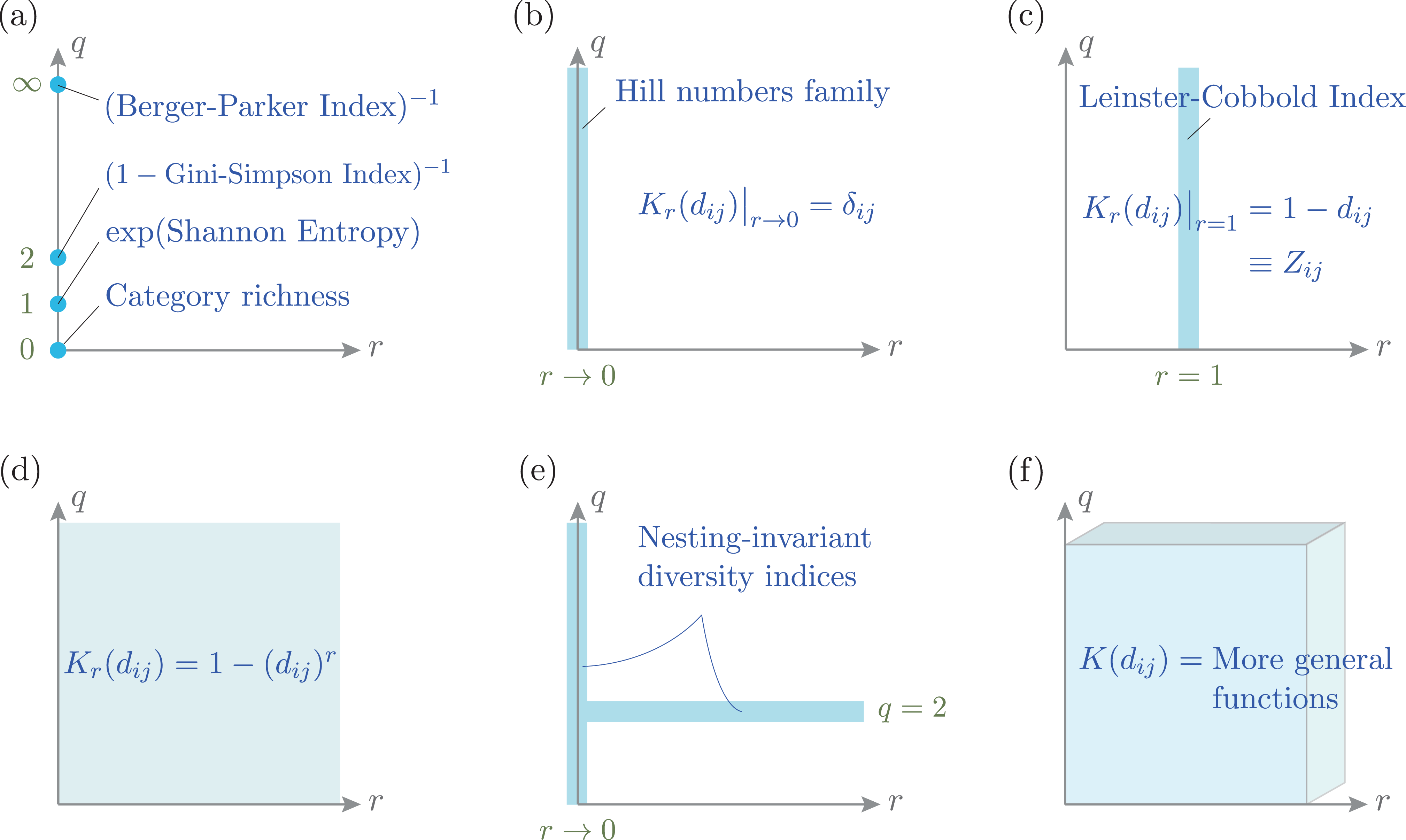}
\vspace{0cm}
\caption{Schematic maps of diversity indices.}
\label{fig:comparison}
\vspace{0.5cm}
\end{figure}

Following Ref.\ \cite{Jost06}, Table \ref{tab:Hill_num} summarises the relations between Hill numbers family and some common diversity/entropy indices, including (Havrda--Charv\'at--Dar\'oczy--)Tsallis entropy \cite{Havrda67,Daroczy70,Tsallis88}, R\'enyi entropy \cite{Renyi61,Jizba04}, Boltzmann--Gibbs--Shannon entropy \cite{Gibbs02,Shannon48,Shannon64}, Simpson concentration \cite{Simpson49}, Gini--Simpson index \cite{Jost06} and Berger--Parker index \cite{Berger70}, which are utilised in various scientific disciplines.
In addition, Fig.\ \ref{fig:comparison} shows a schematic representation of these indices as well as the affinity-based extended diversity indices discussed in the main text.
For all Maps (a)--(f), the vertical and horizontal axes correspond to the order parameter ($q$) of Hill numbers and the scaling parameter ($r$) introduced in this paper, respectively.
Each diversity index in Table \ref{tab:Hill_num} corresponds to a specific point on the spectrum of Hill numbers family (${}^{q}\!D$) indexed by the order parameter $q$, as seen in Maps (a) and (b).
For these diversity indices the effective affinity is given by the Kronecker's delta, $\delta_{ij}$ (see Condition \ref{cond:reduceHill} in Section \ref{subsec:gen.formula}), which can be thought of the $r\to0$ limit of $K_{r}(d_{ij})=1-(d_{ij})^{r}$.
The Leinster--Cobbold diversity index \cite{Leinster12} may be mapped to the $r=1$ case of the same kernel function with identification $Z_{ij}\eq 1-d_{ij}$ (see discussion below Eq.\ (\ref{Tsallis_d})).
Their diversity function has a continuous spectrum with respect to the order parameter $q\in [0,\infty)$, and therefore represented by a straight half-line parallel to the $q$-axis, as shown in Map (c).
If we allow the scaling parameter $r$ in addition to the order parameter $q$ to take an arbitrary positive value, then the allowed region of diversity function covers the entire parameter space spanned by $[0,\infty)\times [0,\infty)$, as shown in Map (d).
In this case, the effect of both the exogenous variables $\{p_{i}\}$ and $\{d_{ij}\}$ on the quantification of diversity can be controlled independently and continuously.
Map (e) represents the allowed region for the nesting-invariant diversity index.
As discussed in the main text, the Nesting Principle is satisfied either when with $q=2$ or when $K(d_{ij})=\delta_{ij}$.
Finally, Map (f) represents the most general forms of affinity-based diversity measures (see the formula (\ref{D-elementary})).
Here the kernel function is not necessary of the form $K_{r}(d_{ij})=1-(d_{ij})^{r}$ but can be of any form as long as it satisfies Conditions \ref{cond:conti}--\ref{cond:even_p}.
The depicted \q{depth} of the parameter space is meant to correspond to other degrees of freedom than $q$ and $r$ characterising the kernel function.

\begin{table}[tbhp]
\centering
\begin{threeparttable}
{\small
\vspace{0.7cm}
\caption{Relations between Hill numbers family and common diversity/entropy indices.}
\label{tab:Hill_num}
\begin{tabular}{ccl}
\addlinespace[7pt]
\toprule
\addlinespace[2pt]
{Order} & {Effective diversity} & {Common index}  \\[2pt]\cline{1-3}
\addlinespace[5pt]
{$q$}\hspace{0.0cm} &$\big(\sum_{i=1}^{n}(p_{i})^q\big)^{\f{1}{1-q}} = \begin{cases}
\, {}^{q}\!D &  \\[0pt]
\, \exp\big(H^{\mathsf{R}}_{q}\big) &  \\[0pt]
\, \exp_{q}\big(H^{\mathsf{T}}_{q}\big)&
\end{cases}$
&$\begin{cases}
\, \text{Hill numbers\,: ${}^{q}\!D$} \\[0pt]
\, \text{R\'enyi entropy\,: $H^{\mathsf{R}}_{q}=\ln {}^{q}\!D=(1-q)^{-1}\ln\sum_{i=1}^{n}(p_{i})^{q}$} \\[0pt]
\, \text{Tsallis entropy\,: $H^{\mathsf{T}}_{q}=\ln_{q}{}^{q}\!D=(1-q)^{-1}\big[\sum_{i=1}^{n}(p_{i})^{q}-1\big]$}
\end{cases}$ \\[5pt]
\addlinespace[3pt]\cdashline{1-3}[2pt/3pt]
\addlinespace[2pt]
$0$\hspace{0.0cm}		&$n$		& {\hspace{2.25mm}Category richness\,: $n={}^{0}\!D$} \\[0pt]
\addlinespace[0pt]
$1$\hspace{0.0cm}		&$\exp\big(-\sum_{i=1}^{n} p_{i}\ln p_{i}\big)=\exp(H_{1})$
&{\hspace{2.25mm}Shannon entropy\,:} $H_{1}=\ln {}^{1}\!D=-\sum_{i=1}^{n} p_{i}\ln p_{i}$ \\[0pt]
\addlinespace[0pt]
$2$\hspace{0.0cm}		& $1\big/\sum_{i=1}^{n}(p_{i})^{2} = \begin{cases}
\, 1\big/I_\mathsf{S} &  \\[0pt]
\, 1\big/(1-I_\mathsf{GS})&
\end{cases}$
&$\begin{cases}
\, \text{Simpson concentration\,: $I_\mathsf{S}=1\big/{}^{2}\!D=\sum_{i=1}^{n}(p_{i})^{2}$} \\[0pt]
\, \text{Gini--Simpson index\,: $I_\mathsf{GS}=1-1\big/{}^{2}\!D=1-\sum_{i=1}^{n}(p_{i})^{2}$}
\end{cases}$ \\[0pt]
\addlinespace[0pt]
$\infty$\hspace{0.0cm}& $1/\max\{p_{1},\,\dots,\,p_{n}\}=1\big/I_\mathsf{BP}$ 	& {\hspace{2.25mm}Berger--Parker index\,: $I_\mathsf{BP}=1\big/{}^{\infty}\!D=\max\{p_{1},\,\dots,\,p_{n}\}$} \\[0pt]
\addlinespace[2pt]
\bottomrule
\end{tabular}
\begin{tablenotes}
\footnotesize
\item 
\textit{Note:} The $q$-exponential function and the $q$-logarithmic function are defined by, respectively,
$\exp_{q}(x)=\big[1+(1-q)x\big]^{\f{1}{1-q}}$ for $1+(1-q)x>0$ and $\ln_{q}(x)=(1-q)^{-1}\big(x^{1-q}-1\big)$ for $x\geq 0$, both for $q\neq 1$.
\end{tablenotes}
\vspace{0.5cm}
}
\end{threeparttable}
\end{table}

\setcounter{figure}{0}
\setcounter{table}{0}
\section{Some Information-Theoretic Properties\label{app:divergence}}

Entropy plays a central role not only in thermodynamics and statistical mechanics but also in many other scientific disciplines, including information theory and statistics.
In this appendix, with future applications in mind, we propose the idea of other affinity-based extended Tsallis type entropies, including the relative entropy (cross entropy), the joint entropy, the conditional entropy and the mutual entropy, each in accordance with the theoretical framework developed in the present paper.
The notations and conventions are as in the main text.\\

We start with the relative entropy.
In the extensive case ($q=1$), it is also known as Kullback--Leibler divergence \cite{Kullback51}, which is regarded as a fundamental quantity as it produces the associated entropy, the mutual information and some probability-distance measures.
Its $q$-generalisation called Tsallis relative entropy is also known in the literature \cite{Tsallis98b,Furuichi04,Furuichi05}.
Here we extend it further so that it can be applied to general affinity-sensitive cases.
Given a pair of marginal probability distributions $\mathbf{P}\coloneqq\{p_{i}\in [0,1]\,|\,\sum_{i\in\cS}p_{i}=1\}$ and $\mathbf{Q}\coloneqq\{q_{i}\in [0,1]\,|\,\sum_{i\in\cS}q_{i}=1\}$\footnote{The probabilities \q{$q_{i}$} are not to be confused with the order/deformation parameter \q{$q$} (without a subscript).} and a pair of dissimilarity variables $\{d_{ij}\in [0,1]\}$ and $\{d'_{ij}\in [0,1]\}$, where $i,\,j\in\cS\coloneqq\{1,\,\dots,\,n\}$ with $n\in\mathbb{Z}^{+}$, the affinity-based extended Tsallis relative entropy of system $\iS\coloneqq\big\{\mathbf{P},\,\{K(d_{ij})\}\big\}$ with respect to system $\iS'\coloneqq\big\{\mathbf{Q},\,\{K(d'_{ij})\}\big\}$ is defined to be
\begin{equation}\label{Tsallis_KL_d}
\mathcal{D}_{q}\big(\iS\parallel\iS'\big)
\coloneqq\begin{cases}
\, \ds -\sum_{i\in\cS}p_{i}\ln_{q}\ko{\f{\sum_{j\in\cS}q_{j}K(d'_{ij})}{\sum_{j\in\cS}p_{j}K(d_{ij})}} & (q\neq 1) \\[4mm]
\, \ds \sum_{i\in\cS}p_{i}\ln\ko{\f{\sum_{j\in\cS}p_{j}K(d_{ij})}{\sum_{j\in\cS}q_{j}K(d'_{ij})}} & (q=1)
\end{cases}\,.
\end{equation}
This can be interpreted as the {effective directed divergence} (\q{effective deviation}) of $\iS$ from $\iS'$ when both the distribution factor and the dissimilarity factor are considered.
Here, $\iS$ may be regarded as the \q{true} system (with \q{true} distribution and dissimilarities) and $\iS'$ as its \q{approximated/modelled} equivalent.\footnote{Although it is possible to consider different kernel functions for the two systems $\iS$ and $\iS'$, here, we consider the same kernel function for simplicity.}
In the zero-affinity limit (i.e.\ $K(d_{ij}),\,K(d'_{ij})\to \delta_{ij}$), the extended Tsallis relative entropy (\ref{Tsallis_KL_d}) reduces to
\begin{align}\label{Tsallis_KL}
{D}_{q}(\mathbf{P}\parallel\mathbf{Q})
=\begin{cases}
\, \ds -\sum_{i\in\cS}p_{i}\ln_{q}\Bigg(\f{q_{i}}{p_{i}}\Bigg)=\f{1}{q-1}\sum_{i\in\cS}\Big[(p_{i})^{q}(q_{i})^{1-q}-1\Big] & (q\neq 1) \\[4mm]
\, \ds \sum_{i\in\cS}p_{i}\ln\Bigg(\f{p_{i}}{q_{i}}\Bigg) & (q=1)
\end{cases}\,,
\end{align}
reproducing Tsallis relative entropy \cite{Tsallis98b,Furuichi04,Furuichi05} including Kullback--Leibler divergence \cite{Kullback51} as its $q=1$ case.
As is well-known, Tsallis relative entropy (\ref{Tsallis_KL}) is always non-negative; it is zero if and only if $\mathbf{P}=\mathbf{Q}$.
This is no longer the case, however, in the presence of nontrivial dissimilarity/affinity factors, as seen in (\ref{Tsallis_KL_d}).
For instance, suppose that a theory failed to consider the dissimilarity factor by assuming $K(d'_{ij})=\delta_{ij}$, whereas the \q{true} system was influenced by the nontrivial effective affinities $\{K(d_{ij})\}$.
Then, the effective deviation of reality from the theory is calculated from (\ref{Tsallis_KL_d}) to be $\mathcal{D}_{q}\big(\iS\parallel\big\{\mathbf{Q},\,\{\delta_{ij}\}\big\}\big)=-\sum_{i\in\cS}p_{i}\ln_{q}\big[1\big/\sum_{j\in\cS}(p_{j}/q_{i})K(d_{ij})\big]$, which does not vanish even if $\mathbf{P}=\mathbf{Q}$.
Due to the same effect of nontrivial $\{K(d_{ij})\}$, other properties such as non-negativity and joint convexity \cite{Furuichi04} neither longer hold in general for the affinity-based extended case.
However, it is shown that a generalised pseudo-additivity relation of the following form is satisfied provided that $\mathbf{P}$ and $\mathbf{Q}$ are independent:
\begin{align}
\mathcal{D}_{q}\big(\iS_{1;\,2}\parallel\iS'_{1;\,2}\big)
=\mathcal{D}_{q}\big(\iS_{1}\parallel\iS'_{1}\big)
+\mathcal{D}_{q}\big(\iS_{2}\parallel\iS'_{2}\big)
+(q-1)\mathcal{D}_{q}\big(\iS_{1}\parallel\iS'_{1}\big)
\mathcal{D}_{q}\big(\iS_{2}\parallel\iS'_{2}\big)\,,
\end{align}
where $\iS_{I}\coloneqq\big\{\{p_{I,i}\},\{K(d_{I,ij})\}\,|\,i,\,j\in\cS_{I}\big\}$ ($I=1,\,2$) and $\iS_{1;\,2}\coloneqq\big\{\{p_{1,i;\,2,j}\},\{K(d_{1,ik;\,2,j\ell})\}\,|\,i,\,k\in\cS_{1},\,j,\,\ell\in\cS_{2}\big\}$ with $p_{1,i;\,2,j}\coloneqq p_{1,i}p_{2,j}$ and $K(d_{1,ik;\,2,j\ell})\coloneqq K(d_{1,ik})K(d_{2,j\ell})$ (or $\rho_{1,ik;\,2,j\ell}=\big((\rho_{1,ik})^{\eta}+(\rho_{2,j\ell})^{\eta}\big)^{1/\eta}$ in terms of the distance variables; see Eq.\ (\ref{d<->r2})), and likewise for $\iS'_{I}$ ($I=1,\,2$) and $\iS'_{1;\,2}$.
As usual, the extended relative entropy is additive if and only if $q=1$.

The case of $\iS'$ with $d'_{ij}=0$ for all $i,\,j\in\cS$, which we denote as $\iS'_{0}$, is of particular interest.
In this case, $\iS'_{0}$ represents a zero-entropy system, and the extended relative entropy of $\iS$ with respect to $\iS'_{0}$ simply measures the extended entropy of $\iS$, i.e.\
\begin{equation}\label{rel_H_D}
\mathcal{D}_{q}\big(\iS\parallel\iS'_{0}\big)=-\cH_{q}(\iS)\,.
\end{equation}
This relation has no analogue in the standard Tsallis case, as it is a consequence of the existence of the new degrees of freedom of affinities.
Together with the fact that $\cH_{q}(\iS)$ is always non-negative, Eq.\ (\ref{rel_H_D}) implies that $\mathcal{D}_{q}\big(\iS\parallel\iS'_{0}\big)$ is always nonpositive and equals zero if and only if $d_{ij}=0$ for all $i,\,j\in\cS$ (i.e.\ when both $\iS$ and $\iS'$ are zero-entropy systems).
This is in contrast to the zero-affinity case, in which ${D}_{q}(\mathbf{P}\parallel\mathbf{Q})$ is always non-negative.\\

Next, the definitions of the joint entropy, the conditional entropy and the mutual entropy in the affinity-based extended theory are in order.
Given a pair of marginal probability distributions $\mathbf{P}\coloneqq\{p_{i}\in [0,1]\,|\,\sum_{i\in\cS}p_{i}=1\}$ and $\mathbf{Q}\coloneqq\{q_{i}\in [0,1]\,|\,\sum_{i\in\cS'}q_{i}=1\}$ and a pair of dissimilarity variables $\{d_{ij}\in [0,1]\,|\,i,\,j\in\cS\}$ and $\{d'_{ij}\in [0,1]\,|\,i,\,j\in\cS'\}$, where $\cS\coloneqq\{1,\,\dots,\,n\}$ and $\cS'\coloneqq\{1,\,\dots,\,n'\}$ with $n,\,n'\in\mathbb{Z}^{+}$, the affinity-based extended Tsallis joint entropy of $\iS\coloneqq\big\{\mathbf{P},\,\{K(d_{ij})\}\big\}$ and $\iS'\coloneqq\big\{\mathbf{Q},\,\{K(d'_{ij})\}\big\}$ is defined to be  
\begin{equation}\label{Tsallis_joint}
\cH_{q}\big(\iS,\iS'\big)
\coloneqq\begin{cases}
\, \ds \sum_{i\in\cS}\sum_{j\in\cS'}p_{ij}\ln_{q}\ko{\f{1}{\sum_{k\in\cS}\sum_{\ell\in\cS'}p_{k\ell}K(d_{ik})K(d'_{j\ell})}} & (q\neq 1) \\[4mm]
\, \ds -\sum_{i\in\cS}\sum_{j\in\cS'}p_{ij}\ln\ko{\sum_{k\in\cS}\sum_{\ell\in\cS'}p_{k\ell}K(d_{ik})K(d'_{j\ell})} & (q=1)
\end{cases}\,,
\end{equation}
where $p_{ij}$ denotes the joint probability of finding the distributions of $\iS$ and $\iS'$ in their $i$-th and $j$-th outcomes, respectively. 
In the zero-affinity limit, Eq.\ (\ref{Tsallis_joint}) reduces to Tsallis joint entropy \cite{Furuichi06}:
\begin{equation}
H_{q}(\mathbf{P},\mathbf{Q})
=\begin{cases}
\, \ds -\sum_{i\in\cS}\sum_{j\in\cS'}(p_{ij})^{q}\ln_{q}p_{ij}
=\f{1}{q-1}\sum_{i\in\cS}\sum_{j\in\cS'}(p_{ij})^{q}\kko{\big(p_{ij}\big)^{1-q}-1} & (q\neq 1) \\[3mm]
\, \ds -\sum_{i\in\cS}\sum_{j\in\cS'}p_{ij}\ln p_{ij} & (q=1)
\end{cases}\,.
\end{equation}
Here, we make some general remarks on the relationship between the extended joint entropy and the effective diversity of the combined system.
First, note that the set of pairs of indices $\{(i,j)\in\cS\times\cS'\}$ in (\ref{Tsallis_joint}) can be uniquely (bijectively) mapped to a new set of indices $\cS_{(2)}\coloneqq\{1,\,\dots,\,nn'\}$.
Hence, the joint entropy (\ref{Tsallis_joint}) can be regarded as representing the extended entropy of a new single system defined on $\cS_{(2)}$ with the new probability distribution $\{\tilde p_{a}\,|\,\sum_{a\in\cS_{(2)}}\tilde p_{a}=1\}$ and the new set of dissimilarity variables $\{\tilde d_{ab}\,|\,a,\,b\in\cS_{(2)}\}$, or of the associated distance variables $\{\tilde \rho_{ab}\,|\,a,\,b\in\cS_{(2)}\}$ (cf.\ Eq.\ (\ref{d<->r2})), so that (\ref{Tsallis_joint}) boils down to exactly the same form as (\ref{Tsallis_d}).
If the mapping is such that $\cS\times\cS'\ni (i,k)\mapsto a\in \cS_{(2)}$ and $\cS\times\cS'\ni (j,\ell)\mapsto b\in \cS_{(2)}$, then the new distance variable is defined in terms of the original distance variables $\{\rho_{ij}\}$ as $\tilde\rho_{ab}\coloneqq\big((\rho_{ik})^{\eta}+(\rho_{j\ell})^{\eta}\big)^{1/\eta}$ for each pair of $(a,b)\in\cS_{(2)}\times\cS_{(2)}$.
It is straightforward to generalise the above argument to multiple $N$ systems case, in which the extended joint entropy of the combined system, $\{\iS_{1},\,\dots,\,\iS_{N}\}$, is given by
\begin{align}\label{Tsallis_joint_N}
&\cH_{q}\big(\iS_{1},\,\dots,\,\iS_{N}\big)
\coloneqq\begin{cases}
\, \ds \sum_{i_{1}\in\cS_{1}}\dots\sum_{i_{N}\in\cS_{N}}p_{i_{1}\dots i_{N}}\ln_{q}\ko{\f{1}{\sum_{j_{1}\in\cS_{1}}\dots\sum_{j_{N}\in\cS_{N}}p_{j_{1}\dots j_{N}}\prod_{I=1}^{N}K(d_{i_{I}j_{I}})}} & (q\neq 1) \\[4mm]
\, \ds -\sum_{i_{1}\in\cS_{1}}\dots\sum_{i_{N}\in\cS_{N}}p_{i_{1}\dots i_{N}}\ln\ko{\sum_{j_{1}\in\cS_{1}}\dots\sum_{j_{N}\in\cS_{N}}p_{j_{1}\dots j_{N}}\prod_{I=1}^{N}K(d_{i_{I}j_{I}})} & (q=1)
\end{cases}\,,\\[4mm]
&\hspace{-8mm}\text{or}\quad \cH_{q}\big(\{\tilde p_{a}\},\{\tilde d_{ab}\}\big)
=\begin{cases}
\, \ds \sum_{a\in\cS_{(N)}}\tilde p_{a}\ln_{q}\ko{\f{1}{\sum_{b\in\cS_{(N)}}\tilde p_{b}K(\tilde d_{ab})}}
=\f{1}{1-q}\sum_{a\in\cS_{(N)}}\tilde p_{a}\kko{\ko{\sum_{b\in\cS_{(N)}}\tilde p_{b}K(\tilde d_{ab})}^{q-1}-1} & (q\neq 1) \\[4mm]
\, \ds -\sum_{a\in\cS_{(N)}}\tilde p_{a}\ln\ko{\sum_{b\in\cS_{(N)}}\tilde p_{b}K(\tilde d_{ab})} & (q=1)
\end{cases}\,,
\end{align}
where $\cS_{(N)}\coloneqq\{1,\,\dots,\,\tilde n\}$ with $\tilde n\coloneqq\prod_{I=1}^{N}n_{I}$ and $n_{I}\coloneqq \abs{\cS_{I}}$, and $\tilde d_{ab}$ is defined such that $K(\tilde d_{ab})=\kappa(\tilde\rho_{ab})$ with $\tilde\rho_{ab}\coloneqq\big(\sum_{I=1}^{N}(\rho_{i_{I}j_{I}})^{\eta}\big)^{1/\eta}$ for each pair of $(a,b)\in\cS_{(N)}\times\cS_{(N)}$ under the mapping of $\bigotimes_{I=1}^{N}\cS_{I}\ni (i_{1},\dots,i_{N})\mapsto a\in \cS_{(N)}$ and $\bigotimes_{I=1}^{N}\cS_{I}\ni (j_{1},\dots,j_{N})\mapsto b\in \cS_{(N)}$.
Subsequently, one can define and calculate the effective diversity of the joint system by using the relation $\exp_{q}(\cH_{q})={}^{q}\!\iD\in [1,\,\tilde n]$.
This interpretability in terms of the effective diversity also ensures some basic properties of the extended joint entropy, including its non-negativity.
It also ensures that the extended joint entropy is nesting-invariant if and only if $q=2$ in the presence of nontrivial affinities between outcomes.
Note also that if all systems ($\iS_{1},\,\dots,\iS_{N}$) are mutually independent, then (\ref{Tsallis_joint_N}) recovers the relation (\ref{N_system}).

In addition, the affinity-based extended Tsallis conditional entropy of $\iS$ given $\iS'$ is defined to be\footnote{Among possible alternatives we adopt this particular definition of the extended Tsallis conditional entropy so that the chain rule (\ref{chain_rule}) holds along with the definition of the extended joint entropy (\ref{Tsallis_joint}).}
\begin{equation}\label{Tsallis_cond}
\cH_{q}\big(\iS\,|\,\iS'\big)
\coloneqq\begin{cases}
\, \ds \sum_{i\in\cS}\sum_{j\in\cS'}p_{ij}\kko{\ln_{q}\ko{\f{1}{\sum_{k\in\cS}\sum_{\ell\in\cS'}p_{k\ell}K(d_{ik})K(d'_{j\ell})}}-\ln_{q}\ko{\f{1}{\sum_{\ell\in\cS'}q_{\ell}K(d'_{j\ell})}}} & (q\neq 1) \\[4mm]
\, \ds -\sum_{i\in\cS}\sum_{j\in\cS'}p_{ij}\ln\ko{\f{\sum_{k\in\cS}\sum_{\ell\in\cS'}p_{k\ell}K(d_{ik})K(d'_{j\ell})}{\sum_{\ell\in\cS'}q_{\ell}K(d'_{j\ell})}} & (q=1)
\end{cases}\,,
\end{equation}
which can be interpreted as the amount of uncertainty remaining about $\iS$ after $\iS'$ has been observed.
In the zero-affinity limit, this reduces to Tsallis conditional entropy \cite{Furuichi06}:
\begin{equation}
H_{q}(\mathbf{P}\,|\,\mathbf{Q})
=\begin{cases}
\, \ds -\sum_{i\in\cS}\sum_{j\in\cS'}(p_{ij})^{q}\ln_{q}p_{i\,|\,j}
=\f{1}{q-1}\sum_{i\in\cS}\sum_{j\in\cS'}(p_{ij})^{q}\kko{\big(p_{i\,|\,j}\big)^{1-q}-1} & (q\neq 1) \\[3mm]
\, \ds -\sum_{i\in\cS}\sum_{j\in\cS'}p_{ij}\ln p_{i\,|\,j} & (q=1)
\end{cases}\,,
\end{equation}
where $p_{i|j}$ denotes the conditional probability of finding the distribution of $\iS$ in its $i$-th outcome, given the distribution of $\iS'$ found in its $j$-th outcome.
Clearly, the extended joint entropy (\ref{Tsallis_joint}) and the extended conditional entropy (\ref{Tsallis_cond}) satisfy the following chain rule:
\begin{equation}\label{chain_rule}
\cH_{q}\big(\iS,\iS'\big)=\cH_{q}\big(\iS\,|\,\iS'\big)+\cH_{q}\big(\iS'\big)\,,
\end{equation}
which generalises the zero-affinity case, $H_{q}(\mathbf{P},\mathbf{Q})=H_{q}(\mathbf{P}\,|\,\mathbf{Q})+H_{q}(\mathbf{Q})$ \cite{Daroczy70,Furuichi06}.
If $\iS$ and $\iS'$ are independent, it can easily be checked that the relation (\ref{chain_rule}) reproduces the pseudo-additivity relation of (\ref{pseudo-additivity}) by noting that $p_{ij}=p_{i}q_{j}$ in the definitions of (\ref{Tsallis_joint}) and (\ref{Tsallis_cond}).
It also follows by induction that the chain rule for multiple ($N$) systems readily holds: $\cH_{q}(\iS_{1},\,\dots,\,\iS_{N})=\sum_{I=1}^{N}\cH_{q}(\iS_{I}\,|\,\iS_{1},\,\dots,\,\iS_{I-1})$.

Moreover, the affinity-based extended Tsallis mutual entropy is defined to be  
\begin{equation}\label{Tsallis_mutual}
\mathcal{I}_{q}\big(\iS;\iS'\big)
\coloneqq\begin{cases}
\, \ds \sum_{i\in\cS}\sum_{j\in\cS'}p_{ij}\kko{\ln_{q}\ko{\f{1}{\sum_{k\in\cS}p_{k}K(d_{ik})}}+\ln_{q}\ko{\f{1}{\sum_{\ell\in\cS'}q_{\ell}K(d'_{j\ell})}}-\ln_{q}\ko{\f{1}{\sum_{k\in\cS}\sum_{\ell\in\cS'}p_{k\ell}K(d_{ik})K(d'_{j\ell})}}} & (q\neq 1) \\[4mm]
\, \ds \sum_{i\in\cS}\sum_{j\in\cS'}p_{ij}\ln\kko{\f{\sum_{k\in\cS}\sum_{\ell\in\cS'}p_{k\ell}K(d_{ik})K(d'_{j\ell})}{\Big(\sum_{k\in\cS}p_{k}K(d_{ik})\Big)\Big(\sum_{\ell\in\cS'}q_{\ell}K(d'_{j\ell})\Big)}} & (q=1)
\end{cases}\,,
\end{equation}
which, in the zero-affinity limit, reduces to Tsallis mutual entropy \cite{Furuichi06}:
\begin{equation}
I_{q}(\mathbf{P};\mathbf{Q})
=\begin{cases}
\, \ds \sum_{i}p_{i}\ln_{q}\big(p_{i}^{-1}\big)+\sum_{i}q_{i}\ln_{q}\big(q_{i}^{-1}\big)-\sum_{i\in\cS}\sum_{j\in\cS'}p_{ij}\ln_{q}\big(p_{ij}^{-1}\big) & (q\neq 1) \\[3mm]
\, \ds \sum_{i\in\cS}\sum_{j\in\cS'}p_{ij}\ln\ko{\f{p_{ij}}{p_{i}q_{j}}} & (q=1)
\end{cases}\,,
\end{equation}
The definition (\ref{Tsallis_mutual}), together with (\ref{Tsallis_joint}) and (\ref{Tsallis_cond}), indicates that the following relations hold:
\begin{equation}
\mathcal{I}_{q}\big(\iS;\iS'\big)
=\cH_{q}(\iS)-\cH_{q}\big(\iS\,|\,\iS'\big)
=\cH_{q}(\iS')-\cH_{q}\big(\iS'\,|\,\iS\big)
=\cH_{q}(\iS)+\cH_{q}(\iS')-\cH_{q}\big(\iS,\iS'\big)\,,
\end{equation}
which generalise the equivalent relations in the zero-affinity case.
Evidently, the mutual entropy vanishes in the case of $\iS'=\iS'_{0}$ (i.e.\ $d'_{ij}=0$ for all $i,\,j\in\cS'$), when the affinity-based extended entropy (\ref{Tsallis_d}), the conditional entropy (\ref{Tsallis_cond}) and the joint entropy (\ref{Tsallis_joint}) are related as $\cH_{q}(\iS)=\cH_{q}\big(\iS\,|\,\iS'_{0}\big)=\cH_{q}\big(\iS,\iS'_{0}\big)$.
Again, this feature has no analogue in the standard Tsallis type entropies.
For instance, based on the usual (implicit) assumption of $K(d_{ij})=K(d'_{ij})=\delta_{ij}$ (i.e.\ the standard Tsallis/Shannon case), the entropy, $H_{q}(\mathbf{P})$, is greater than the conditional entropy, $H_{q}(\mathbf{P}\,|\,\mathbf{Q})$, unless the two distributions $\mathbf{P}$ and $\mathbf{Q}$ are independent.
By contrast, here, the difference between $\cH_{q}(\iS)$ and $\cH_{q}\big(\iS\,|\,\iS'\big)$ can be arbitrarily close to zero even if $\mathbf{P}$ and $\mathbf{Q}$ are not independent, provided that the underlying between-outcome dissimilarities of $\iS'$ can also be close to zero.

\setcounter{figure}{0}
\setcounter{table}{0}
\section{Implications for Biodiversity\label{app:biodiversity}}

The quantification of biodiversity is fundamental for biological, ecological and environmental studies, including various disciplines such as population dynamics, ecological interactions and conservation biology.
A number of diversity indices have been proposed in the literature; however, these often lead to results that are inconsistent and sometimes contradictory to one another.
As such, there has not been a universally accepted framework for measuring, comparing and partitioning diversity.
Herein, we propose a systematic method to evaluate biodiversity that is applicable not only to species diversity but also to other types of biodiversity, including genetic diversity and ecosystem diversity.
The formulation is based on the affinity-based extended diversity measure and the Nesting Principle developed in the main text.
In particular, it provides a solution to the long-standing problem of how to partition the overall biodiversity (\q{$\gamma$-diversity}) into the within-population component (\q{$\alpha$-diversity}) and the between-population component (\q{$\beta$-diversity}) in a mathematically and biologically consistent manner.
In what follows, the notations and conventions are as in the main text.

First, let us review some essential aspects of diversity quantification in the biological context, specifically in terms of species diversity.
It is often characterised by the number of species (\q{species richness}), the relative abundances of individuals within each species (\q{species abundance}) and the between-species dissimilarities.
As an illustrative example, consider the following three biological populations, each containing six individuals but differ in their species compositions: $\mathsf{P}_{1}=\{\mathsf{a},\,\mathsf{b},\,\mathsf{c},\,\mathsf{d},\,\mathsf{e},\,\mathsf{f}\}$, $\mathsf{P}_{2}=\{\mathsf{a},\,\mathsf{a},\,\mathsf{b},\,\mathsf{b},\,\mathsf{c},\,\mathsf{c}\}$ and $\mathsf{P}_{3}=\{\mathsf{a},\,\mathsf{a},\,\mathsf{a},\,\mathsf{a},\,\mathsf{b},\,\mathsf{c}\}$, where each letter denotes maximally dissimilar species.
Denoting the diversity of $\mathsf{P}_{I}$ as $D_{I}$ ($I = 1,\,2,\,\dots$), common diversity measures \cite{MacArthur65,Hill73} conclude that $D_{1}>D_{2}$ and $D_{1}>D_{3}$, as the species richness is higher in $\mathsf{P}_{1}$ than in $\mathsf{P}_{2}$ or $\mathsf{P}_{3}$.
Common diversity measures \cite{Hill73} also conclude that $D_{2}>D_{3}$ because the individuals are more evenly distributed among species in $\mathsf{P}_{2}$ than in $\mathsf{P}_{3}$.
Other measures, which give as much importance to rare species as to the common ones, may conclude that $D_{2}=D_{3}$ because the species richness is equivalent between the two populations.
Consider also the following population with yet another composition: $\mathsf{P}_{4}=\{\mathsf{a},\,\textrm{A},\,\mathscr{A},\,a,\,\mathcal{A},\,\mathbb{A}\}$.
It contains six distinct species that are relatively quite similar (i.e.\ all species possess variations of the same character). In this case, common diversity measures \cite{Rao82,Solow93,Leinster12} would conclude that $D_{1}>D_{4}$ because dissimilarities among the species are greater in $\mathsf{P}_{1}$ than in $\mathsf{P}_{4}$, whereas other measures, which give equal importance to similar and dissimilar species, may conclude that $D_{1}=D_{4}$ because both populations have the same species richness.
Thus, the quantification of diversity is influenced by two \q{objective} (or \q{exogenous}) aspects---species abundance \cite{MacArthur65,Hill73,Jost06,Simpson49,Berger70} and between-species dissimilarity (in terms of some quantifiable character) \cite{Rao82,Solow93,Leinster12}\footnote{Here the \q{species richness} aspect has been incorporated in the definition of \q{species abundance} (see Footnote \ref{fnlabel} in the main text).}---and two \q{subjective} (or \q{endogenous}) aspects---the weight given to species abundance \cite{Hill73,Jost06,Simpson49,Berger70}, which is well-acknowledged in the literature, and that given to between-species dissimilarity, which has attained little attention so far.

Here we note that the quantification of dissimilarity itself is not a trivial task.
In the above illustration, we implicitly assumed that all individuals categorised in \q{$\mathsf{a}$} are identical entities, i.e.\ $d_{\mathsf{a}\mathsf{a}}\equiv 0$.
However, a closer look at these individuals may reveal that they are, in fact, rather different in their characteristics (such as, e.g., shape, colour, weight, or surface pattern) even if they are of the same species \q{$\mathsf{a}$}.
Then, from a finer-grained point of view, for instance, the population $\mathsf{P}_{3}=\{\mathsf{a},\,\mathsf{a},\,\mathsf{a},\,\mathsf{a},\,\mathsf{b},\,\mathsf{c}\}$ may have been better recognised as $\mathsf{P}'_{3}=\{\mathsf{a},\,\mathsf{a}^{\prime},\,\mathsf{a}^{\prime\prime},\,\mathsf{a}^{\prime\prime\prime},\,\mathsf{b},\,\mathsf{c}\}$ with nontrivial between-individual dissimilarities (i.e.\ $d_{\mathsf{a}\mathsf{a}^{\prime}}>0$ and the same for other pairs of \q{$\mathsf{a}$}s)---if not the differences may be as prominent as those seen among the species in $\mathsf{P}_{4}$.
Thus, with sufficiently high resolution, one can no longer ignore the dissimilarities among those once with lower resolution regarded as identical.
This is a situation where the between-individual ($\mathsf{a},\,\mathsf{a}^{\prime},\,\mathsf{a}^{\prime\prime},\,\dots$) diversity is nested in the between-species ($\mathsf{a},\,\mathsf{b},\,\mathsf{c},\,\dots$) diversity, which can further be nested into the between-population ($\mathsf{P}_{1},\,\mathsf{P}_{2},\,\mathsf{P}_{3},\,\dots$) diversity, the between-community diversity and so on.
Alternatively, the between-species diversity could also be nested according to taxonomy, i.e.\ into the between-genus diversity, the between-family diversity and so on.
In this way, there naturally arises the notion of what can be referred to as the \q{nested diversities}.
Here the nests are not necessarily formed hierarchically in the order of proximity or similarity among the constituents but can be implemented in an arbitrarily intertwined manner.

\begin{figure}
\centering
\includegraphics[width=0.98\linewidth,clip]{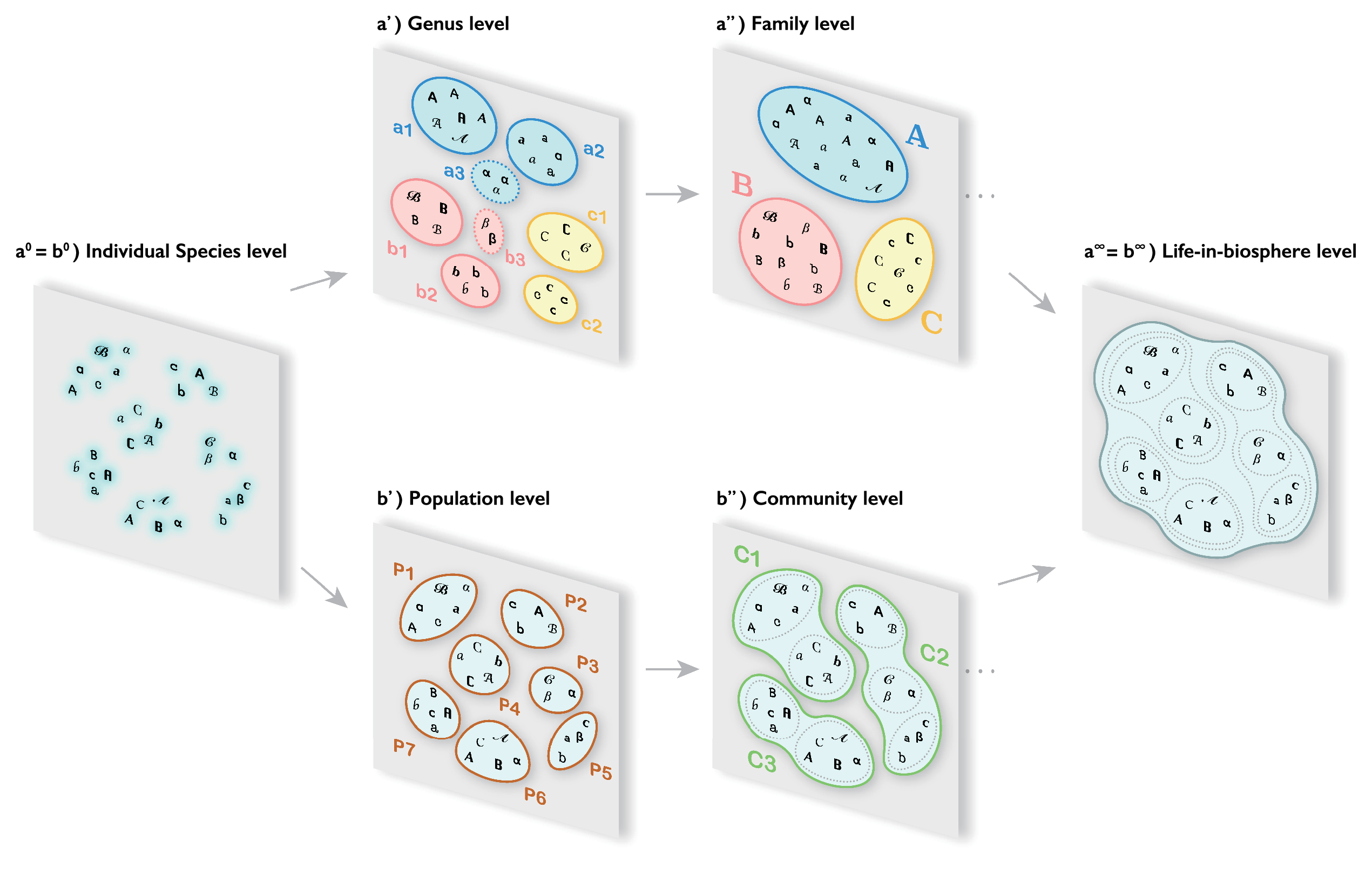}
\caption{Coarse-graining flow of nested and grouped diversities.
The set of effective quantities, $\{P_{I}\}$ and $\{\bar{d}_{IJ}\}$, can be computed at each level of nesting in each of the two coarse-graining procedures, $\mathbf{a}$ and $\mathbf{b}$.
Each grouped and coloured region represents a coarse-grained category at each level of nesting.
For instance, if the nesting structure is determined by taxonomic hierarchy, then the flow can be observed from $\mathbf{a}^{0}$) \q{Species} to $\mathbf{a}^{\prime}$) \q{Genus} (a1--a3, b1--b3 and c1--c2) to $\mathbf{a}^{\prime\prime}$) \q{Family} (A, B and C) level, and so forth, up to $\mathbf{a}^{\infty}$) \q{Life} at the most coarse-grained level.
Alternatively, if the nesting structure is determined by geographic proximity, then the flow can be observed from $\mathbf{b}^{0}$) \q{Individual} to $\mathbf{b}^{\prime}$) \q{Population} (P1--P6) to $\mathbf{b}^{\prime\prime}$) \q{Community} (C1--C3) level, and so forth, up to $\mathbf{b}^{\infty}$) the \q{Biosphere} level at the ultimate limit.
The Nesting Principle ensures that the diversity function $\iD(\{P_{I}\},\{\bar{d}_{IJ}\})$ produces the same diversity index value at any cross section of these flows.}
\label{fig:nesting_chain}
\end{figure}

A mathematically consistent framework enabling the implementation of this notion of {nested diversities} has already been developed in the main text through the Nesting Principle.
Let us consider a population in which there are a total of $n$ species labelled by $i\in \cS=\{1,\,\dots,\,n\}$, each with relative abundance $p_{i}\in [0,1]$ such that $\sum_{i\in\cS} p_{i}=1$.
Let us also consider its partitioning into $N$ groups labelled by $I\in\cN=\{1,\,\dots,\,N\}$ such that $\cS=\bigcup_{I\in\cN}\cS_{I}$ with $\abs{\cS_{I}}\in\mathbb{Z}^{+}$ and $\sum_{I\in\cN}\abs{\cS_{I}}=n$.
Then, the Nesting Principle states that the diversity computed at the \q{bare} (without-nesting) level, $\iD_\mathrm{bare} (\{p_{i}\},\{d_{ij}\})$, and that computed at the aggregate meta-level, $\iD_\mathrm{nest} (\{P_{I}\},\{\bar{d}_{IJ}\})$, give the same value.
Here, $\{P_{I}\}$ and $\{\bar{d}_{IJ}\}$ represent the effective equivalents of $\{p_{i}\}$ and $\{d_{ij}\}$, respectively, defined at the meta-level biological categories labelled by $\{I\}$.
In other words, the Nesting Principle ensures that the set of effective quantities, $\{P_{I}\}$ and $\{\bar{d}_{IJ}\}$ (and $\{\iD_{I}\}$), can be consistently defined and computed for an arbitrary nesting or grouping of the constituents so that they produce the same value of the overall diversity index, $\iD$, which should be the case for any consistent biodiversity evaluation.
Fig.\ \ref{fig:nesting_chain} demonstrates two types of nesting structures, or procedures of coarse-graining, which can develop because of either a hierarchy in taxonomy (flow $\mathbf{a}$) or a geographical proximity (flow $\mathbf{b}$), for example. Starting with the same initial system $\mathbf{a}^{0}=\mathbf{b}^{0}$ (i.e.\ the collective set of all elementary species described by $\{p_{i}\}$ and $\{d_{ij}\}$), both flows of the coarse-graining procedure eventually reach the same final outcome $\mathbf{a}^{\infty}=\mathbf{b}^{\infty}$ (i.e.\ the set of all living things in the biosphere as the ultimate single \q{category} with $P = 1$).
Essentially, the Nesting Principle dictates that the diversity function $\iD(\{P_{I}\},\{\bar{d}_{IJ}\})$ results in the same diversity index value at any cross-section of these flows: i.e.\ $\iD(\mathbf{a}^{0})=\iD(\mathbf{a}^{\prime})=\iD(\mathbf{a}^{\prime\prime})=\iD(\mathbf{a}^{\infty})$, which is also equal to $\iD(\mathbf{b}^{0})=\iD(\mathbf{b}^{\prime})=\iD(\mathbf{b}^{\prime\prime})=\iD(\mathbf{b}^{\infty})$.

The affinity-sensitive diversity measure that is conceptualised as the effective number of species, satisfying Conditions \ref{cond:conti}--\ref{cond:even_p} of Section \ref{subsec:gen.formula} and also compatible with the Nesting Principle, is given by the formula (\ref{coarse-grainedD2});
here we collect relevant formulae for readers' easy reference:
\begin{equation}\label{formula}
\iD\big(\{P_{I}\},\{\bar{d}_{IJ}\},\{\iD_{I}\}\big)
=\Bigg[1-\sum_{I,\,J\in\cN}\f{P_{I}}{P}\f{P_{J}}{P}\big(\bar{d}_{IJ}\big)^{r}\Bigg]^{-1^{\vphantom{\f{1}{1}}}}\,,
\end{equation}
\vspace{-5mm}
\begin{subequations}
  \begin{empheq}[left={\quad\text{where}\quad }\empheqlbrace]{align}
     ~~&P_{I}(\{p_{i}\}) =\sum_{i\in\cS_{I}}p_{i}\,,\quad 
     \sum_{I\in\cN}P_{I}=1\,,\label{formula_a}\\[0pt]
     ~~&\bar{d}_{IJ}(\{p_{i}\},\{d_{ij}\}) =\Bigg[\sum_{i\in\cS_{I}}\sum_{j\in\cS_{J}}\hf{p_{i}\vphantom{p_{j}}}{P_{I}}\hf{p_{j}}{P_{J}}\,\big(d_{ij}\big)^{r}\Bigg]^{1/r}\,,\label{formula_b}\\[0pt]
      ~~&\iD_{I}(\{p_{i}\},\{d_{ij}\}) =\Bigg[1-\sum_{i,\,j\in\cS_{I}}\f{p_{i}}{P_{I}}\f{p_{j}}{P_{I}}\big(d_{ij}\big)^{r}\Bigg]^{-1}\,.\label{formula_c}
  \end{empheq}
\end{subequations}

Given that, we now consider the problem of how to partition the overall diversity ($\gamma$) into the within-population component ($\alpha$) and the between-population component ($\beta$) while meeting various technical or ecological constraints.
A number of methods have been previously proposed to address this problem \cite{Whittaker60,McArthur66,Wilson84,Lande96,Veech02,Jost07,Tuomisto10a,Tuomisto10b,Veech10,Baselga10}; however, no definitive quantitative formulation is available in the literature.
Below we show how the affinity-based extended framework developed in the main text provides a solution to this long-standing problem.
Let $\alpha_{I}$ and $\alpha_{J}$ denote the diversity of sub-populations $I$ and $J$, respectively, and $\gamma$ denote the diversity associated with the whole population, i.e.\ the union of $I$ and $J$, denoted as $I\cup J$.
Let $\beta$ denote the $\beta$-diversity defined in relation to these $\alpha$- and $\gamma$-components.
Then, the following correspondence between the population-biological concepts and the building blocks of the affinity-based diversity index can be naturally established for the current two-population ($I$ and $J$) setup:
\begin{equation}\label{correspondence}
\alpha_{I}~\leftrightarrow~\iD_{I}\,,\quad
\alpha_{J}~\leftrightarrow~\iD_{J}\,,\quad
\gamma~\leftrightarrow~\iD_{I\cup J}\quad
\text{and}\quad
\beta_{IJ}~\leftrightarrow~ (\bar{d}_{IJ})^{r}\,,
\end{equation}
where $\iD_{X}$ denotes the diversity index (\ref{formula}) computed for (sub-)population $X$.
Subsequently, the following formula is obtained by solving (\ref{formula}) for the $\beta$-component:
\begin{equation}\label{def:beta}
\beta=\f{1}{2}\kko{ \f{P_{I}}{P_{J}}\ko{\f{1}{\alpha_{I}}-\f{1}{\gamma}}+\f{P_{J}}{P_{I}}\ko{\f{1}{\alpha_{J}}-\f{1}{\gamma}}  }
+\ko{1-\f{1}{\gamma}}\,.
\end{equation}
This concludes that the $\alpha$-, $\beta$- and $\gamma$-diversity components are related neither by the simple additive (i.e.\ symbolically, \q{$\alpha+\beta=\gamma$}) \cite{McArthur66,Lande96,Veech02,Veech10} nor by the multiplicative (i.e.\ \q{$\alpha\times\beta=\gamma$}) \cite{Whittaker60,Jost07,Veech10,Baselga10} relation, but are related in the integrated manner as in (\ref{def:beta}).
It also indicates that the common assumption that $\alpha$-diversity does not exceed $\gamma$-diversity \cite{Lande96} does not hold in general because either one of the $\alpha$-diversities (say, $\alpha_{I}$) can exceed $\gamma$.
This is possible simply because $\gamma$ is determined not only by $\alpha_{I}$, $\alpha_{J}$ and $\beta$ but also by the ratio of relative abundances $P_{J}/P_{I}$.
Specifically, $\alpha_{I}>\gamma$ is possible if $\alpha_{J}<\gamma$ is satisfied and $P_{J}/P_{I}$ is sufficiently large.

In this connection, population biologists have also often asked the question of what is the effective number of sub-populations present in a meta-population.
Given the above formulation and identification of diversity components, the answer to this question depends on our assumptions as to how to characterise a \q{typical} sub-population here.
For instance, if we simply assume that the meta-population can be regarded as composed of $M$ (finite number) sub-populations of equal size ($1/M$), equal $\alpha$-diversity ($\bar{\alpha}$), and with equal $\beta$-diversity ($\bar{\beta}$) between any two different sub-populations, then the formula (\ref{formula}) leads to
\begin{equation}\label{partition}
M = \f{\bar{\alpha}^{-1}+\bar{\beta}-1}{\gamma^{-1}+\bar{\beta}-1}\,.
\end{equation}
The interpretation of this relation is straightforward.
On the one hand, if we set $\bar{\beta}=1$ (i.e.\ all the sub-populations are maximally dissimilar to each other), then the relation (\ref{partition}) simply becomes the condition for the replication principle \cite{Hill73,Jost06}: $\gamma=M\bar{\alpha}$.
If moreover $\bar{\alpha}=1$ (i.e.\ each population is composed of a single species), then this relation further reduces to the maximum diversity case: $\gamma=M$.
On the other hand, if we set $\bar{\beta}=0$ and also consider the elementary case ($\bar{\alpha}=1$), then we obtain $\gamma=1$, corresponding to the minimum diversity case.\\

To obtain better sense of how the diversity formulae (\ref{formula}) and (\ref{formula_a})--(\ref{formula_c}) can be applied to the biodiversity evaluation in terms of the above-identified $\alpha$-, $\beta$- and $\gamma$-diversities, we provide some numerical demonstrations.
We employ the following specific set (\q{fruit basket}) of $n=9$ items:
\begin{equation}\label{tot.basket}
\cB_\mathrm{all}=\{\mathsf{a},\,\mathsf{a}^{\prime},\,\mathsf{b},\,\mathsf{b}^{\prime},\,\mathsf{b}^{\prime\prime},\,\mathsf{c},\,\mathsf{c},\,\mathsf{c},\,\mathsf{c}\}\,,
\end{equation}
where the dissimilarity among the items are displayed in Table \ref{tab:dissimilarity_fruit}.
Here each symbolised fruit item is given a specific name for ease of imagination (which is totally hypothetical, to be sure).
For instance, the dissimilarity between species $\mathsf{a}$ (\q{apple}) and $\mathsf{b}$ (\q{orange}) can be read off from the table as $d_{\mathsf{a}\mathsf{a}^{\prime}}=0.88$.
The symbol $\varepsilon$ represents a very small positive number that is negligible at the species level.
It can be checked that the given set of dissimilarities satisfies the triangle inequality: $d_{ij}+d_{jk}\geq d_{ik}$ for all $i,\,j,\,k\in\cB_\mathrm{all}$.
The relative abundance of each fruit item is given by:
\begin{equation}
p_{\mathsf{a}}=p_{\mathsf{a}^{\prime}}=p_{\mathsf{b}}=p_{\mathsf{b}^{\prime}}=p_{\mathsf{b}^{\prime\prime}}=\f{1}{9}
~~\text{and}~~
p_{c}=\f{4}{9}\,.\no
\end{equation}

\begin{table}[t]
\centering
\vspace{-0.5cm}
\caption{Dissimilarities $\{d_{ij}\}$ among \q{fruit species} (hypothetical data).}
\label{tab:dissimilarity_fruit}
{\small
\begin{tabular}{cc;{2pt/2pt}ccc;{2pt/2pt}ccccll}
$\mathsf{a}$ & $\mathsf{a}^{\prime}$   & $\mathsf{b}$    & $\mathsf{b}^{\prime}$   & $\mathsf{b}^{\prime\prime}$ & $\mathsf{c}$ & $\mathsf{c}$ & $\mathsf{c}$ & $\mathsf{c}$ & & \\ \cline{1-9}
0 & 0.28 & \underline{0.88} & 0.90  & 0.89 & 0.92 & 0.92 & 0.92 & \multicolumn{1}{c|}{0.92} & $\mathsf{a}$   & (apple)      \\
& 0    & 0.86 & 0.83 & 0.85 & 0.95 & 0.95 & 0.95 & \multicolumn{1}{c|}{0.95} & $\mathsf{a}^{\prime}$  & (pear)       \\ \cdashline{3-11}[2pt/2pt]
&     \multicolumn{1}{c}{} & 0    & 0.12 & 0.13 & 0.96 & 0.96 & 0.96 & \multicolumn{1}{c|}{0.96} & $\mathsf{b}$   & (orange)     \\
&     \multicolumn{1}{c}{} &      & 0    & 0.19 & 0.97 & 0.97 & 0.97 & \multicolumn{1}{c|}{0.97} & $\mathsf{b}^{\prime}$  & (grapefruit) \\
&     \multicolumn{1}{c}{} &      &      & 0    & 0.95 & 0.95 & 0.95 & \multicolumn{1}{c|}{0.95} & $\mathsf{b}^{\prime\prime}$ & (lemon)      \\ \cdashline{6-11}[2pt/2pt]
&     \multicolumn{1}{c}{} &      &      & \multicolumn{1}{c}{}     & 0    & $\varepsilon$    & $\varepsilon$    & \multicolumn{1}{c|}{$\varepsilon$}    & $\mathsf{c}$  & (strawberry) \\
&     \multicolumn{1}{c}{} &      &      & \multicolumn{1}{c}{}     &      & 0    & $\varepsilon$    & \multicolumn{1}{c|}{$\varepsilon$}    & $\mathsf{c}$  & (strawberry) \\
&     \multicolumn{1}{c}{} &      &      & \multicolumn{1}{c}{}     &      &      & 0    & \multicolumn{1}{c|}{$\varepsilon$}    & $\mathsf{c}$  & (strawberry) \\
&     \multicolumn{1}{c}{} &      &      & \multicolumn{1}{c}{}     &      &      &      & \multicolumn{1}{c|}{0}    & $\mathsf{c}$  & (strawberry)
\end{tabular}
}
\vspace{0.5cm}
\end{table}

\noindent
For the moment, let us set $r=1$ for simplicity.
First, by applying the diversity formula (\ref{formula_c}), the diversity of the whole basket is computed as
\begin{equation}\label{iD_tot}
\gamma\coloneqq\iD_\mathrm{tot}
=\Bigg[1-\sum_{i,\,j\in\cB_\mathrm{all}}p_{i}p_{j}d_{ij}\Bigg]^{-1}~\approx~\underline{2.60}\,.
\end{equation}
This result can be intuitively understood as follows:
there are roughly three kinds of fruits in $\cB_\mathrm{all}$, that are: $\mathsf{a}$'s, $\mathsf{b}$'s and $\mathsf{c}$'s (as indicated by the dotted separation lines in Table \ref{tab:dissimilarity_fruit}), while the similarity among them reduces the effective variety, leading to the diversity index less than three.

Now, let us decompose the whole set into the following three subsets according to the species similarity:
\begin{equation}\label{basket1-3}
\cB_{1}=\{\mathsf{a},\,\mathsf{a}^{\prime}\}\,,\quad
\cB_{2}=\{\mathsf{b},\,\mathsf{b}^{\prime},\,\mathsf{b}^{\prime\prime}\}\,,\quad
\cB_{3}=\{\mathsf{c},\,\mathsf{c},\,\mathsf{c},\,\mathsf{c}\}\no
\end{equation}
so that $\cB_\mathrm{all}=\cB_{1}\cup\cB_{2}\cup\cB_{3}$ (disjoint union).
The effective number of species in each basket is computed by applying the formula (\ref{formula_c}) to each basket $I=1,\,2,\,3$; namely, $\alpha_{I}\coloneqq \iD_{I}=\big[1-\sum_{i,\,j\in\cB_{I}}(p_{i}/P_{I})(p_{j}/P_{I})d_{ij}\big]^{-1}$.
The between-basket $\beta$-diversity, $\beta_{IJ}\coloneqq\bar{d}_{IJ}$, can also be computed by the formula (\ref{formula_b}).
For example, the diversity of $\cB_{2}$, denoted as $\alpha_{2}$, and the effective dissimilarity between $\cB_{1}$ and $\cB_{2}$, denoted as $\beta_{12}$, are computed to be, respectively,
\begin{align}
\alpha_{2}
&=\bigg[1-2 \bigg(\f{p_{\mathsf{b}}}{P_{2}}\f{p_{\mathsf{b}^{\prime}}}{P_{2}}d_{\mathsf{b}\mathsf{b}^{\prime}}
+\f{p_{\mathsf{b}}}{P_{2}}\f{p_{\mathsf{b}^{\prime\prime}}}{P_{2}}d_{\mathsf{b}\mathsf{b}^{\prime\prime}}
+\f{p_{\mathsf{b}^{\prime}}}{P_{2}}\f{p_{\mathsf{b}^{\prime\prime}}}{P_{2}}d_{\mathsf{b}^{\prime}\mathsf{b}^{\prime\prime}}\bigg)\bigg]^{-1}
~\approx~\underline{1.11}\,,\no\\[6pt]
\beta_{12}
&=\f{p_{\mathsf{a}}}{P_{1}}\f{p_{\mathsf{b}}}{P_{2}}d_{\mathsf{a}\mathsf{b}}
+\f{p_{\mathsf{a}}}{P_{1}}\f{p_{\mathsf{b}^{\prime}}}{P_{2}}d_{\mathsf{a}\mathsf{b}^{\prime}}
+\f{p_{\mathsf{a}}}{P_{1}}\f{p_{\mathsf{b}^{\prime\prime}}}{P_{2}}d_{\mathsf{a}\mathsf{b}^{\prime\prime}}
+\f{p_{\mathsf{a}^{\prime}}}{P_{1}}\f{p_{\mathsf{b}}}{P_{2}}d_{\mathsf{a}^{\prime}\mathsf{b}}
+\f{p_{\mathsf{a}^{\prime}}}{P_{1}}\f{p_{\mathsf{b}^{\prime}}}{P_{2}}d_{\mathsf{a}^{\prime}\mathsf{b}^{\prime}}
+\f{p_{\mathsf{a}^{\prime}}}{P_{1}}\f{p_{\mathsf{b}^{\prime\prime}}}{P_{2}}d_{\mathsf{a}^{\prime}\mathsf{b}^{\prime\prime}}
~\approx~\underline{0.868}\,.\no
\end{align}
These results are shown in Table \ref{tab:3basket}, along with the relative abundances $\{P_{I}\}$ for each basket that is obtained from (\ref{formula_a}).
Note, in particular, that the effective dissimilarity within $\cB_{1}$, that is $\beta_{11}$, is no longer zero but has a finite value ($\approx 0.140$) due to the heterogeneity of species within the basket.
The same holds for $\beta_{22}$.
By contrast, $\beta_{33}$ is zero since the basket $\cB_{3}$ is homogeneous in species composition.
Note also that $\alpha_{I}=(1-\beta_{II})^{-1}$ holds for each $I$, which manifests the equivalence relation (\ref{d_II}).

\begin{table}[h]
\centering
\vspace{5mm}
\caption{The $\beta$-diversity indices $\{\beta_{IJ}\}$, the relative abundances $\{P_{I}\}$ and the $\alpha$-diversity indices $\{\alpha_{I}\}$ for Baskets $\cB_{1}$--$\cB_{3}$.}
\label{tab:3basket}
{\small
\begin{tabular}{ccclll}
$\cB_{1}$ & $\cB_{2}$ & $\cB_{3}$ & \hspace{1cm} &\hspace{2cm}  &  \\ \cline{1-3}
0.140 & \underline{0.868} & \multicolumn{1}{c|}{0.935} & $\cB_{1}$ & $P_{1}=2/9$ & $\alpha_{1}\approx 1.16$ \\
& 0.098 & \multicolumn{1}{c|}{0.960} & $\cB_{2}$ & $P_{2}=1/3$ & $\alpha_{2}\approx \underline{1.11}$ \\
$\boxed{\beta_{IJ}}$ &  & \multicolumn{1}{c|}{0} & $\cB_{3}$ & $P_{3}=4/9$ & $\alpha_{3}=1$
\end{tabular}
}
\vspace{5mm}
\end{table}

\noindent
Given that, let us recompute the diversity of the whole set $\cB_\mathrm{all}$, this time using the coarse-grained, basket-level variables as the building blocks of diversity.
Using the formula (\ref{formula}), this leads to:
\begin{equation}\label{iD_tot2}
\iD_\mathrm{tot,\,nest}
=\Bigg[1-\sum_{I,\,J\in\{1,\,2,\,3\}}P_{I}P_{J}\beta_{IJ}\Bigg]^{-1}~\approx~\underline{2.60}\,,
\end{equation}
which, by virtue of the Nesting Principle, agrees with $\gamma$ in (\ref{iD_tot}) computed at the most elementary species level.

In fact, as guaranteed by the Nesting Principle, this conclusion holds for an arbitrary decomposition of the initial full basket.
In particular, the coarse-graining procedure to reach the basket-level does not necessarily be performed according to the species similarity (like \q{\textsl{Citrus Basket}} or \q{\textsl{Malus Basket}} as in the previous decomposition).
To illustrate this, take another set of baskets:
\begin{equation}\label{basket1-4}
\cB_{1^{\prime}}=\{\mathsf{a},\,\mathsf{b},\,\mathsf{c}\}\,,\quad
\cB_{2^{\prime}}=\{\mathsf{a}^{\prime},\,\mathsf{b}^{\prime},\,\mathsf{c}\}\,,\quad
\cB_{3^{\prime}}=\{\mathsf{b}^{\prime\prime},\,\mathsf{c}\}\,,\quad
\cB_{4^{\prime}}=\{\mathsf{c}\}\no
\end{equation}
so that, again, $\cB_\mathrm{all}=\cB_{1^{\prime}}\cup\cB_{2^{\prime}}\cup\cB_{3^{\prime}}\cup\cB_{4^{\prime}}$ (disjoint union).
Table \ref{tab:4basket} summarises the values of $\beta_{I^{\prime}J^{\prime}}\coloneqq\bar{d}_{I^{\prime}J^{\prime}}$, $\{P_{I^{\prime}}\}$ and $\alpha_{I^{\prime}}\coloneqq \iD_{I^{\prime}}$ for the baskets $\cB_{I^{\prime}}$ ($I^{\prime}=1^{\prime}\,,\dots,\,4^{\prime}$), computed using the data in Table \ref{tab:dissimilarity_fruit} and the diversity formulae (\ref{formula_a})--(\ref{formula_c}).
Then, by plugging the new set of coarse-grained variables listed in Table \ref{tab:4basket} in the formula (\ref{formula}), the effective number of species in $\cB_\mathrm{all}$ can be computed as follows:
\begin{equation}\label{iD_tot3}
\iD_\mathrm{tot,\,nest^{\prime}}
=\Bigg[1-\sum_{I^{\prime},\,J^{\prime}\in\{1^{\prime},\,2^{\prime},\,3^{\prime},\,4^{\prime}\}}P_{I^{\prime}}P_{J^{\prime}}\beta_{I^{\prime}J^{\prime}}\Bigg]^{-1}~\approx~\underline{2.60}\,,
\end{equation}
which reproduces the diversity index of (\ref{iD_tot}).
These results, yielding $\mathrm{(\ref{iD_tot})=(\ref{iD_tot2})=(\ref{iD_tot3})}$, provide simple manifestation of the Nesting Principle for a closed system (this time composed of fruit items).

\begin{table}[h]
\centering
\vspace{5mm}
\caption{The $\beta$-diversity indices $\{\beta_{I^{\prime}J^{\prime}}\}$, the relative abundances $\{P_{I^{\prime}}\}$ and the $\alpha$-diversity indices $\{\alpha_{I^{\prime}}\}$ for Baskets $\cB_{1^{\prime}}$--$\cB_{4^{\prime}}$.}
\label{tab:4basket}
{\small
\begin{tabular}{cccclll}
$\cB_{1^{\prime}}$ & $\cB_{2^{\prime}}$ & $\cB_{3^{\prime}}$ & $\cB_{4^{\prime}}$ & \hspace{1cm} & \hspace{2cm} &  \\ \cline{1-4}
0.613 & 0.662 & 0.642 & \multicolumn{1}{c|}{0.627} & $\cB_{1^{\prime}}$ & $P_{1^{\prime}}=1/3$ & $\alpha_{1^{\prime}}\approx 2.59$ \\
& 0.611 & 0.652 & \multicolumn{1}{c|}{0.640} & $\cB_{2^{\prime}}$ & $P_{2^{\prime}}=1/3$ & $\alpha_{2^{\prime}}\approx 2.57$ \\
&  & 0.475 & \multicolumn{1}{c|}{0.475} & $\cB_{3^{\prime}}$ & $P_{3^{\prime}}=2/9$ & $\alpha_{3^{\prime}}\approx 1.90$ \\
$\boxed{\beta_{I^{\prime}J^{\prime}}}$ &  &  & \multicolumn{1}{c|}{0} & $\cB_{4^{\prime}}$ & $P_{4^{\prime}}=1/9$ & $\alpha_{4^{\prime}}=1$
\end{tabular}
}
\vspace{5mm}
\end{table}

\begin{figure}[tb]
\centering
\includegraphics[width=0.7\linewidth,clip]{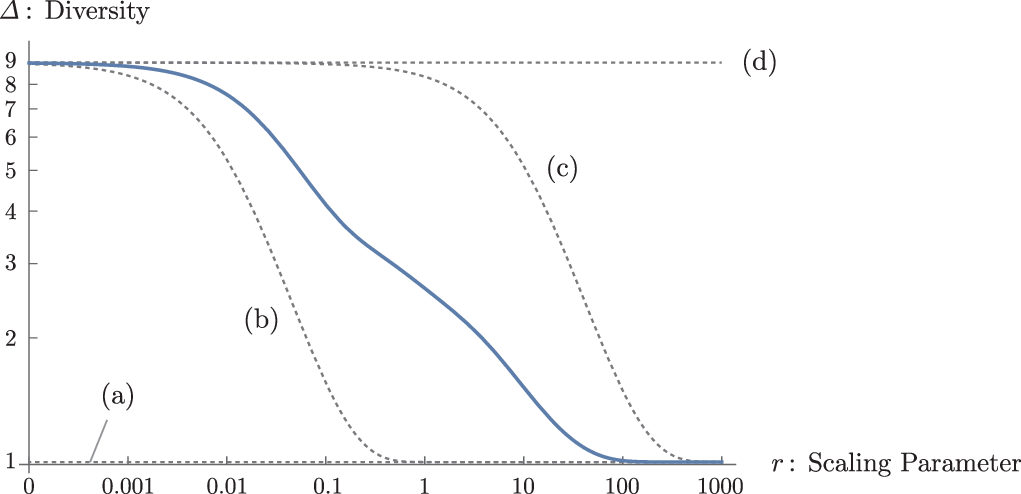}
\vspace{0.5cm}
\caption{The profile of diversity function $\iD^{(r)}(\{p_{i}\},\{d_{ij}\})$ based on the data of $\{p_{i}\}$ and $\{d_{ij}\}$ given in Table \ref{tab:dissimilarity_fruit} (with $\ep=10^{-6}$).
The log--log plot of $\iD^{(r)}$ as a function of the scaling parameter $r$ is shown.
As references, diversity profiles corresponding to different values of $d_{i\neq j}=\bar{d}=\mathrm{const.}$ are also shown by dotted lines: $\mathrm{(a)}$ $\bar{d}=0$, $\mathrm{(b)}$ $\bar{d}=10^{-4}$, $\mathrm{(c)}$ $\bar{d}=0.99$ and $\mathrm{(d)}$ $\bar{d}=1$.}
\label{fig:graph_scaling}
\vspace{0.5cm}
\end{figure}

Although we have considered the case of $r=1$, for practical applications, the scaling parameter can be adjusted in response to particular needs and the context of diversity evaluation.
As discussed in the main text, in the limit $r\to 0$, the diversity index (\ref{formula}) reduces to Hill numbers \cite{Hill73}.
If $r\in (0,1)$, then the diversity index assigns a greater weight to relatively similar species pairs, whereas if $r\in(1,\infty)$, then it assigns a greater weight to relatively dissimilar species pairs, with the limiting case $r\to \infty$ reflecting only the prevalence of the most dissimilar species pairs.
Fig.\ \ref{fig:graph_scaling} shows the effect of $r$ on the diversity quantification, in which the diversity index is computed for the particular case of $\cB_\mathrm{all}$ as given in Table \ref{tab:dissimilarity_fruit}.
The profile of diversity continuously varies with $r$.
In particular, $\iD_{r\to 0}=n=9$ and $\iD_{r\to\infty}=1$, both regardless of the actual relative abundances of individual items (corresponding to the two \q{extreme observer} cases).\\

Finally, apart from the issue of diversity partitioning, we make some remarks on the relations between the Nesting Principle and other principles or properties known in the literature.
As already noted in Footnote \ref{fn:nesting}, the principle of splitting/merging invariance discussed in Section \ref{subsec:gen.formula} can be regarded as a special case (i.e.\ weaker version) of the Nesting Principle.
Specifically, it is obtained by setting $d_{ij}=0$ for $i,\,j\in\cS_{1}=\{1,2\}$ while keeping $\{d_{k\ell}\}$ for $k,\,\ell\in\cS_{2}=\{3,\,\dots,\,n\}$ general.
In addition, the replication principle \cite{Hill73,Jost06}, whose $N=2$ case is usually referred to as the \q{doubling property}, and the modularity principle \cite{Leinster12} are also special cases of the Nesting Principle.
The modularity principle states that if a population is partitioned into $N$ sub-populations ($\cS=\bigcup_{I\in\cN}\cS_{I}$), with species in different sub-populations being maximally dissimilar ($d_{ij}=1-\delta_{ij}$ for all $i\in\cS_{I}$ and all $j\in\cS_{J}$ with $I\neq J$), then the diversity of the whole population is determined by
\begin{equation}\label{modularity}
\iD=\Bigg[ \sum_{I\in\cN}\ko{\f{P_{I}}{P}}^{q}(\iD_{I})^{1-q}\Bigg]^{\f{1}{1-q}}\,.
\end{equation}
Evidently, both (\ref{aggregateD}) and (\ref{coarse-grainedD}) reduce to (\ref{modularity}) under the above-stated condition.
The replication principle states that if, moreover, these $N$ sub-populations are of equal size ($P_{I}/P=1/N=\mathrm{const.}$) and equal diversity ($\iD_{I}=\bar{\alpha}=\mathrm{const.}$), then the diversity of the whole population is given by $N$ times the diversity of a sub-population, i.e.\ $\iD=N\bar{\alpha}$.
These assumptions about the modularity principle and the replication principle are, however, unrealistic in practical applications since in reality species in different sub-populations will generally be more or less similar in some aspects.

\bibliographystyle{elsarticle-num} 

\providecommand{\noopsort}[1]{}\providecommand{\singleletter}[1]{#1}%

\end{document}